\documentclass[review]{elsarticle}

\usepackage{lineno,hyperref}
\biboptions{sort&compress}
\usepackage{amsmath}
\usepackage{bm}
\usepackage{algorithm}
\usepackage{algcompatible}
\usepackage{algpseudocode}
\usepackage{float}
\usepackage{graphicx,subfigure}
\usepackage{epstopdf}
\usepackage{amsmath,bm}
\usepackage{booktabs}
\usepackage{mathtools}
\usepackage{array}
\usepackage{tabularx}
\usepackage{caption}
\usepackage{wasysym}
\usepackage{filecontents,lipsum}
\usepackage{subfigure}
\usepackage{xcolor}
\usepackage{hhline}
\usepackage{color}
\usepackage{adjustbox}
\usepackage{dblfloatfix}
\usepackage{amssymb}
\newcommand{\ignore}[1]{}
\DeclarePairedDelimiter\ceil{\lceil}{\rceil}

\journal{Computer Networks}

\begin{document}

\begin{frontmatter}

\title{Robust QoT Assured Resource Allocation in Shared Backup Path Protection Based \ignore{Elastic Optical Networks}EONs}


\author{Venkatesh Chebolu}
\author{Sadananda Behera}
\author{Goutam Das,\textit{ Member, IEEE}}
\address{Indian Institute of Technology, Kharagpur,India}




%

\begin{abstract}
Survivability is mission-critical for elastic optical networks (EONs) as they are expected to carry an enormous amount of data. In this paper, we consider the problem of designing shared backup path protection (SBPP) based EON that facilitates the minimum quality-of-transmission (QoT) assured allocation against physical layer impairments (PLIs) under any single link/shared risk link group (SRLG) failure for static and dynamic traffic scenarios. In general, the effect of PLIs on lightpath varies based on the location of failure of a link as it introduces different active working and backup paths. 
To address these issues in the design of SBPP EON, we formulate a mixed integer linear programming (MILP) based robust optimization framework for static traffic with the objective of minimizing overall fragmentation. In this process, we use the efficient bitloading technique for spectrum allocation for the first time in survivable EONs. In addition, we propose a novel SBPP-impairment aware (SBPP-IA) algorithm considering the limitations of MILP for larger networks. For this purpose, we introduce a novel sorting technique named most congested working-least congested backup first (MCW-LCBF) to sort the given set of static requests. Next, we employ our SBPP-IA algorithm for dynamic traffic scenario and compare it with existing algorithms in terms of different QoT parameters. We demonstrated through simulations that our study provides around 40\% more QoT guaranteed requests compared to existing ones.
\end{abstract}


\end{frontmatter}


	\section{Introduction} \label{introduction}
	Elastic optical networks (EONs) have emerged as a potential candidate to replace the existing wavelength division multiplexed (WDM) networks due to their superior attributes in terms of huge data transportation and high spectrum efficiency \cite{chatterjee2017fragmentation}. In general, optical networks suffer from unexpected link and shared risk link group (SRLG) failures which may cause the failure of multiple lightpaths resulting in immense loss of data and revenue. In order to overcome this, it is imperative to design a survivable EON that enables the normal operation of the network during link/SRLG failures. Several EON based protection techniques, namely, span restoration, p-cycles, dedicated protection, shared backup path protection (SBPP) etc. are proposed in the literature \cite{shen2016survivable}. Among all, SBPP has been widely adopted as it offers major benefits such as effective protection capacity sharing, fast recovery, and less complexity \cite{shen2016survivable}. Therefore, in this work, we focus on SBPP based EON. 
	
	Routing and spectrum assignment (RSA) is an integral part of network planning wherein a request is assigned a route and wavelength(s). In a transparent optical network, the transmission quality of a lightpath is affected by various physical layer impairments (PLIs). Note that, existing works related to SBPP based EON \cite{shen2014optimal, anoh2017efficient, guo2016routing, cai2016multicast, lu2018cost, papanikolaou2017joint, yin2016shared, wang2018availability, chen2014spectrum, wu2017energy, tang2017mixed, moghaddam2018routing, chen2014spectrum2, wang2015distance, lourencco2017algorithm, luo2019leveraging, hsu2019spectrum} ignored the effects of PLIs during RSA which might lead to the degraded quality-of-transmission (QoT) of lightpath signals in terms of bit error rate (BER) both in working and backup paths. On the other hand, few works \cite{fontinele2017efficient,chatterjee2021impairment,zhou2020link,zhao2019holding,behera2018effect,behera2019impairment,wang2019load,zhao2015nonlinear,krishnamurthy2020physical} in literature addressed the problem of assuring minimum QoT for lightpath signals against PLIs but were implemented only for unprotected networks. This results in severe problems during the link/SRLG failures as mentioned before. Therefore, considering the practical scenarios, it is indispensable to design RSA in EON which protects the network against link/SRLG failures as well as ensures the minimum QoT for lightpath signals in both working and backup paths. Thus, in this paper, we design QoT guaranteed RSA in presence of PLIs in SBPP based EON against single link/SRLG failures.

	It should be noted that, QoT assured SBPP EON cannot be designed easily by just combining existing studies related to SBPP EONs and QoT guaranteed unprotected EONs.The problem of ensuring QoT in survivable EON is fundamentally different and challenging than that of unprotected EONs due to the following reasons. In unprotected EONs, ensuring QoT guaranteed allocation for any request is the function of allocations of existing requests in the network as interference due to PLIs on any frequency slot (FS)  depends on allocations. When it comes to survivable EONs, interference scenario is completely different. This is because, if we consider a particular link/SRLG failure in the network, working paths of existing requests containing the failed link/SRLG are deactivated and corresponding backup paths are activated. From this, we can observe that, set of requests whose working paths are active and the set of requests whose backup paths are active in the network varies with the failed link/SRLG. Therefore, different interference scenarios exist in survivable EON for different link/SRLG failures. Thus, unlike unprotected EONs, ensuring QoT guaranteed allocation for any request in survivable EONs is not only a function of allocations of existing requests but also a function of failed link/SRLG.  
	Since we do not know which link/SRLG of the network will fail beforehand, allocation should be such that whichever link/SRLG fails QoT is guaranteed in all working and backup paths. For this purpose, we need to consider each possible allocation in the network corresponding to each single link/SRLG failure case before establishing working and backup path of each request. Considering above mentioned issues, we propose a \textit{robust optimization} based RSA in this study wherein allocation is done in working and backup path of each request by taking the worst case interference scenario among all interferences corresponding to each single link/SRLG failure case.

	\vspace{0.1cm}
	
	The rest of the paper is organized as follows. Section \ref{State Of the Art} and \ref{Motivation} describe the state of the art and challenges, motivation, contributions of our problem respectively. Section \ref{MILP}, \ref{heuristic} and \ref{SRLG} present the MILP formulation, the heuristic algorithm and the SRLG extension respectively. Section \ref{results} provides simulation results for performance evaluation while Section \ref{conclusion} concludes the paper.
	
	\section{State Of the Art} \label{State Of the Art}
	In this section, we present the existing studies related to RSA design in SBPP based EON and QoT guaranteed unprotected EON. Under the static traffic scenario, Shen et al. \cite{shen2014optimal} developed integer linear programming (ILP) models for both SBPP and 1+1 path protection based EONs to analyze and compare their performance. They also examined the impact of both transponder tunability and bandwidth squeezed restoration techniques. Authors in \cite{anoh2017efficient} presented a survivable hybrid protection lightpath (HybPL) algorithm to minimize spectrum utilization in the network. In this algorithm, resource availability and power consumption together decide whether to provide dedicated or shared path protection. In \cite{guo2016routing}, authors focused on RSA design of survivable EON against dual link failures for various sharing capabilities of backup lightpaths. For this purpose, they developed ILP and heuristics with the goal of minimizing total number of FSs used. Cai et al. formulated mixed integer linear programming (MILP) and proposed a heuristic algorithm for routing, modulation level, and spectrum assignment (RMLSA) problem for multicast capable EON with shared protection \cite{cai2016multicast}. Authors in \cite{lu2018cost} considered integrated multilayer protection problem in IP over EONs against single link or router failure at any time. To address this problem, they presented an ILP model and proposed efficient algorithms. Various other works related to shared protection based EON under static traffic are presented in  \cite{papanikolaou2017joint,yin2016shared,wang2018availability,chen2014spectrum,wu2017energy,tang2017mixed,moghaddam2018routing}. 
	
	Considering the dynamic traffic scenario, authors in \cite{chen2014spectrum2} presented minimum free spectrum-block consumption algorithm to solve the RSA problem for SBPP EON with the objective of reducing total spectrum usage while satisfying the joint failure probability threshold. Authors in \cite{wang2015distance} focused on distance adaptive RSA problem for EON with SBPP for which they proposed spectrum window plane based heuristic algorithms. Moreover, authors in this study considered differentiated shareable FS cost to share the backup FSs more efficiently. In addition, significant research has been carried out regarding dynamic SBPP EON in \cite{lourencco2017algorithm,luo2019leveraging,hsu2019spectrum}. However, all the above mentioned studies did not consider the effect of PLIs in RSA design of SBPP EON.
	
	With respect to PLI aware studies, authors in \cite{fontinele2017efficient,chatterjee2021impairment,zhou2020link,zhao2019holding,behera2018effect,behera2019impairment,wang2019load,zhao2015nonlinear,krishnamurthy2020physical} considered various impairments such as in-band crosstalk, nonlinear impairments, filtering effects and amplified spontaneous emission noise etc. in RSA to ensure minimum QoT for all requests. But, the authors in these studies did not extend their work to design survivable EONs.
	
	As mentioned in the introduction, existing works related to SBPP EON ignored the effect of PLIs in RSA design and the studies where PLIs were taken into account in RSA design have not considered the protection against link/SRLG failures. Therefore, it is imperative to design a robust RSA in EON to ensure both protection against link/SRLG failures as well as minimum QoT in working and backup lightpaths against PLIs.
	
	\begin{figure*}[t!]
		\hspace{-0.28cm}	
		\subfigure[]{\includegraphics[trim={-1cm 9cm 20cm 0cm},height=0.4\textwidth, width=.35\textwidth]{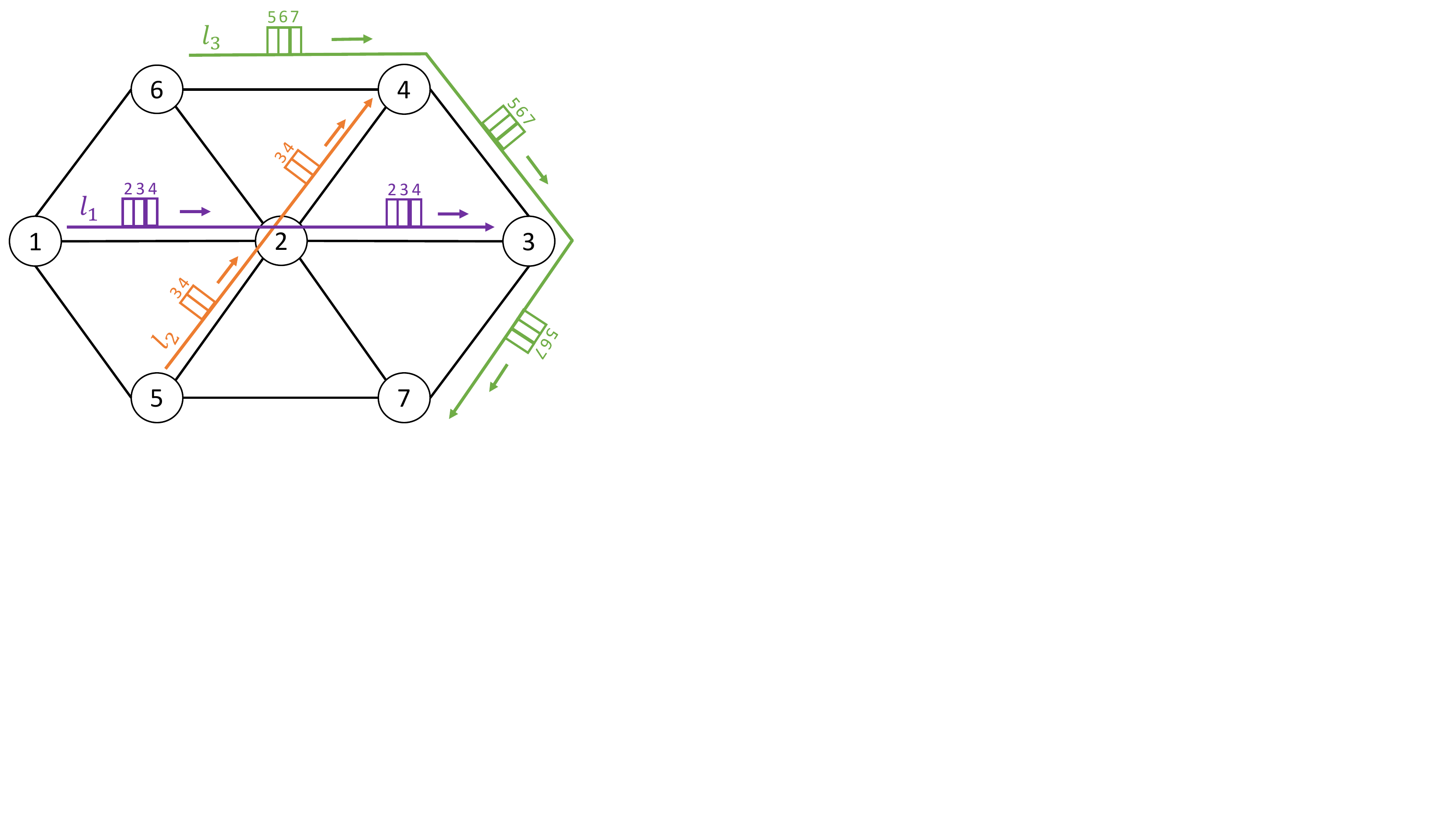} \label{no_failure}} 
		\hspace{-0.5cm}
		\subfigure[]{\includegraphics[trim={-0.5cm 9cm 19.5cm 0cm},height=0.4\textwidth, width=.35\textwidth]{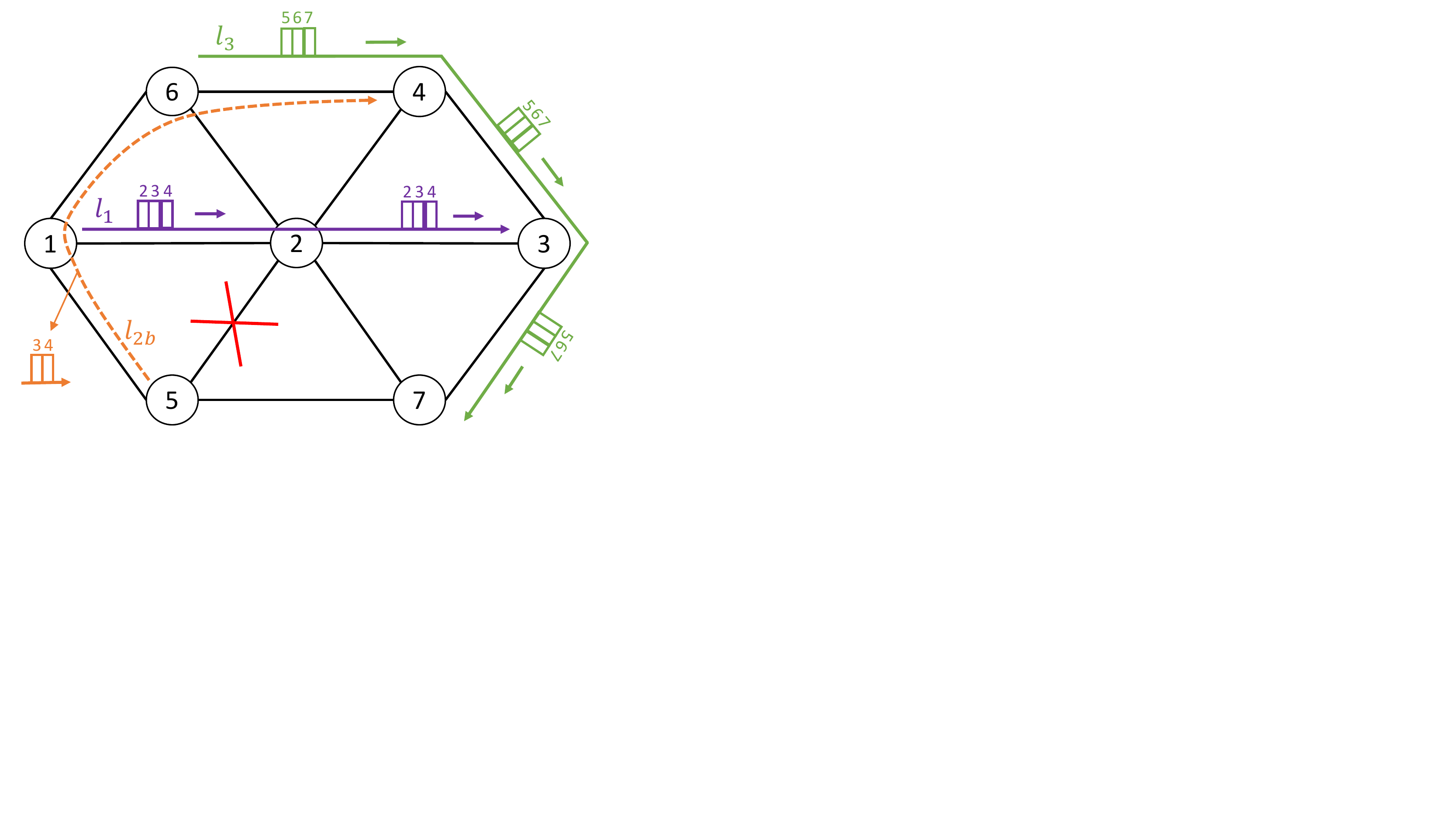}	\label{link1_failure}}
		\hspace{-0.8cm}
		\subfigure[]{\includegraphics[trim={-1.5cm 10.5cm 19.5cm -0.5cm},height=0.38\textwidth, width=.35\textwidth]{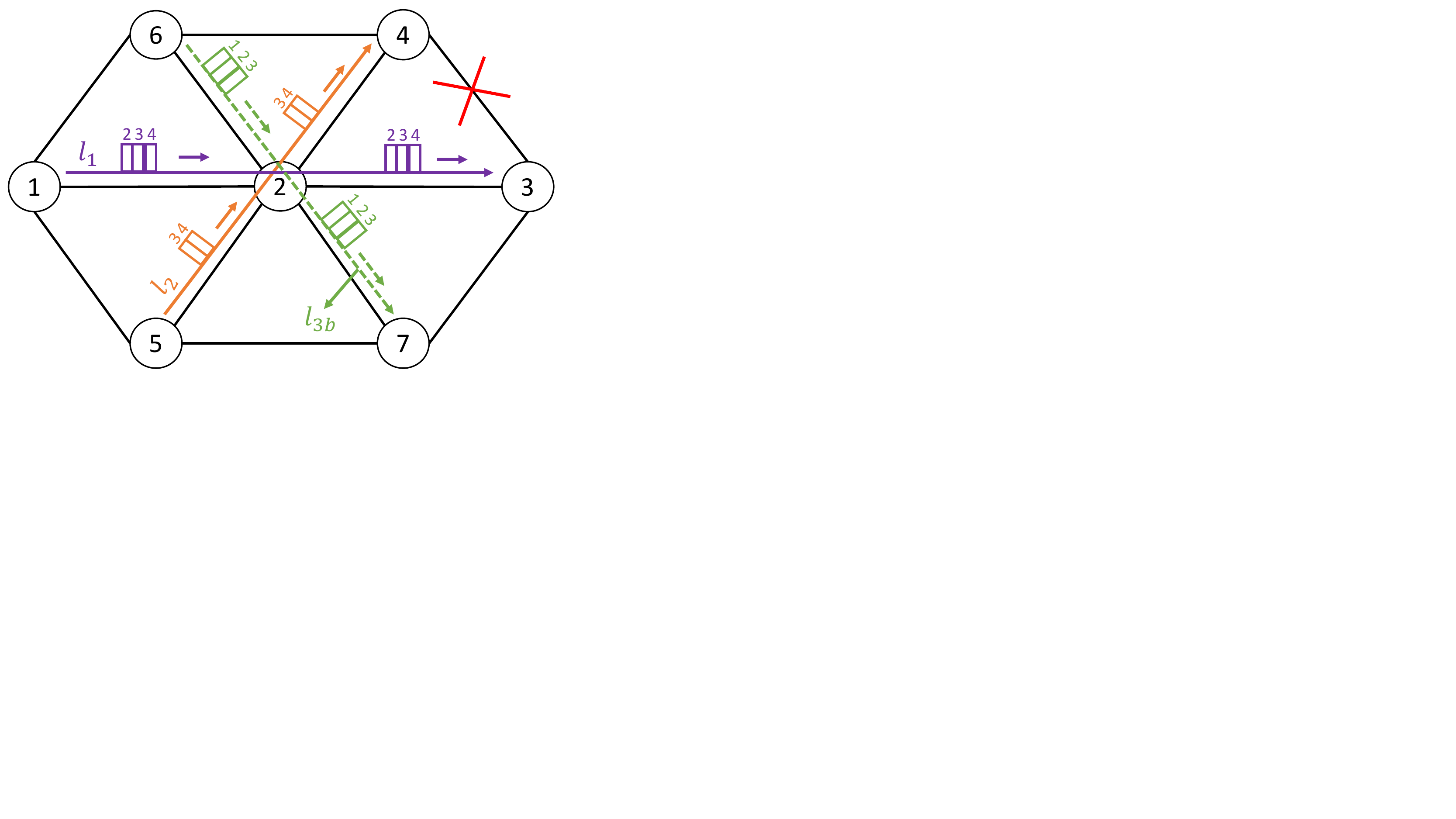}	\label{link2_failure}} 
		\caption{Effect of in-band crosstalk on FS `3' at XC `2' under different link failure conditions. (a) No link failure (b) 5-2 link failure (c) 4-3 link failure.}
		\label{example_fig}
	\end{figure*}

	\begin{figure}[h] 
		    \hspace{0.5cm}
		\includegraphics[trim={0cm 4cm 23.5cm 0.5cm}, height=0.6\textwidth, width=0.5\textwidth]{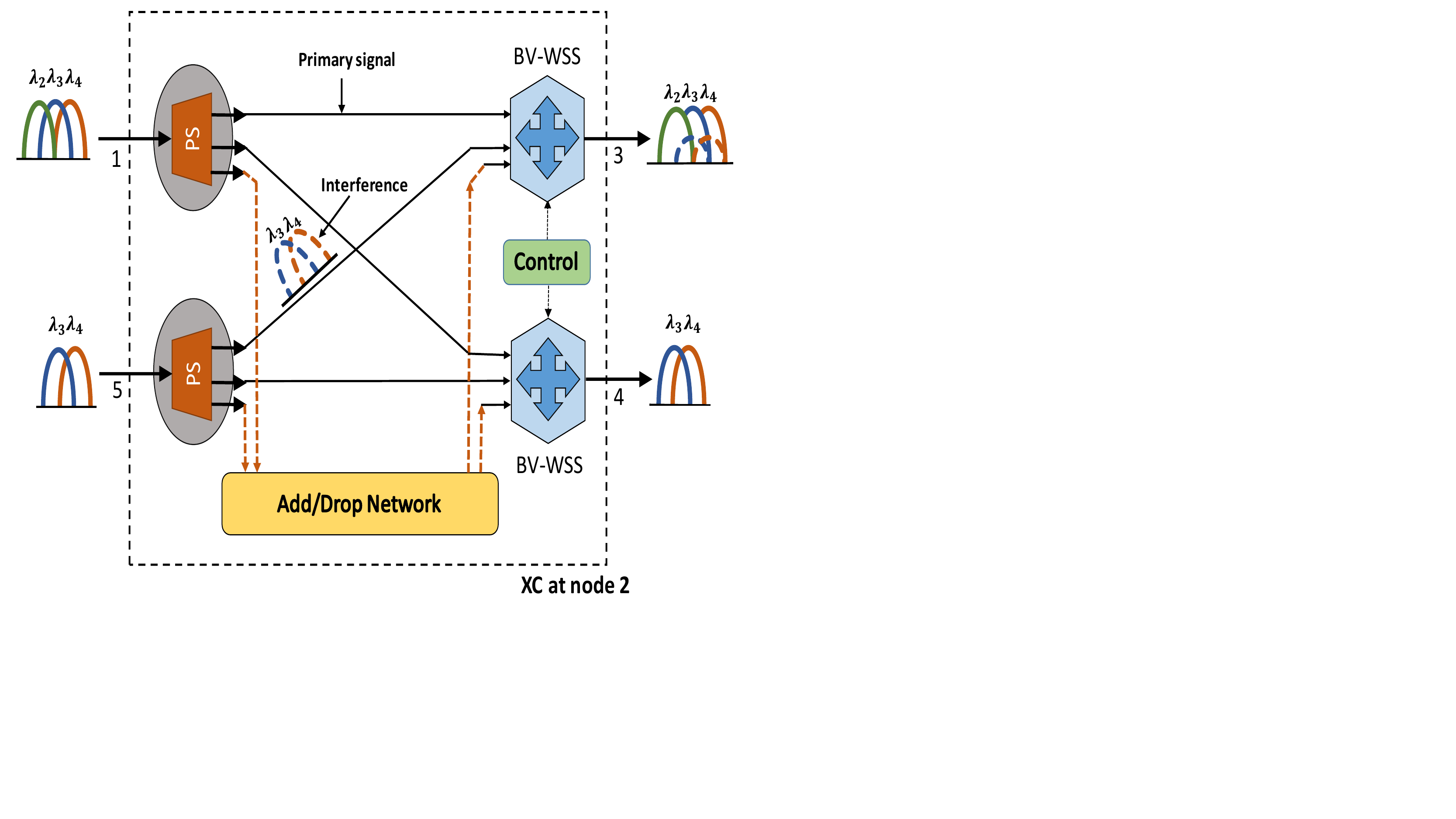}\label{crosstalk3}	
		\vspace{-0.5cm}
		\caption{Formation of in-band crosstalk in cross connect (XC) at node 2 ; PS: passive splitter; BV-WSS: bandwidth variable wavelength selective switch.}	
		\label{inband_crosstalk}     
	\end{figure}
	
	\section{Challenges, motivation and contributions} \label{Motivation}
	In this section, we describe the challenges and motivation of our considered problem. In this regard, we discuss the concept of PLIs first. PLIs such as shot noise, amplified spontaneous emission (ASE) noise, crosstalk (in-band and out-of-band), nonlinear impairments, filtering effects and beat noises due to coherent reception, etc., degrade the quality of lightpath signal and limit its optical reach \cite{behera2019impairment}.
	Among these, in-band crosstalk (IXT) is severely coupled with the RSA framework \cite{behera2018effect}. Therefore, in this work, we target to ensure the minimum QoT for working and backup light paths signals by considering the IXT as a major source of interference along with ASE and beating noise terms due to coherent reception. However, as shown in \cite{behera2019impairment}, our formulation can be easily extended for nonlinear impairments and filter narrowing effects. In general, the generation of IXT in any optical network is due to the non-ideal nature of the switch used in cross connect (XC). When two or more lightpath signals co-propagate through the same XC, subcarrier of each signal gets interference from the subcarriers of other signals having same frequency. This effect will be more if the signal propagates through multiple XCs in its path. Now, we illustrate the challenges associated with IXT to ensure the minimum QoT in working and backup paths of each request in SBPP based EON with a suitable example.

	\subsection{Challenges to ensure QoT in SBPP EON} \label{challenges}
	We consider a 7-node network as shown in Fig. \ref{example_fig} in which each link is bi-directional. Each rectangular box in Fig. \ref{example_fig} denotes the FS or subcarrier that is being used by the request. A request $r_i$ is described as $r_i(s,d)$ where, \textit{s} and \textit{d} stands for source and destination, respectively. We consider 3 requests ${r}_{1}(1,3)$, ${r}_{2}(5,4)$ and ${r}_{3}(6,7)$. Provisioning of working paths for these 3 requests is shown in Fig. \ref{no_failure} represented by ${l}_{1}$,${l}_{2}$ and ${l}_{3}$. Here, we analyze the effect of IXT on FS `3' in lightpath ${l}_{1}$ at XC `2' under different link failure conditions. For this purpose, first we assume that there is no failure in the network as shown in Fig. \ref{no_failure}. In this case, FS `3' in ${l}_{1}$ experiences IXT interference from the FS `3' present in lightpath ${l}_{2}$ at XC `2' as shown in Fig. \ref{inband_crosstalk}. Note that, $\lambda_2$,$\lambda_3$ and $\lambda_4$ shown in Fig. \ref{inband_crosstalk} denotes FS 2,3 and 4 respectively. Next, we consider that link 5-2 is failed as shown in Fig. \ref{link1_failure}. Due to this, working path ${l}_{2}$ of request ${r}_{2}$ is affected which is rerouted through the backup path denoted as ${l}_{2b}$. In this case, FS `3' in ${l}_{1}$ does not get interference from any other lightpaths at XC `2'. Similarly, for 4-3 link failure, working path ${l}_{3}$ of request ${r}_{3}$ is affected and switch over is carried out to its backup path ${l}_{3b}$ as shown in Fig. \ref{link2_failure}. Here, FS `3' in ${l}_{1}$ gets interference at XC `2' from lightpaths ${l}_{2}$ and ${l}_{3b}$ which are using FS `3'. From the above discussion, we made the following observations. Firstly, set of existing requests whose working paths are active ($ R_w $) and the set of existing requests whose backup paths are active ($ R_b $) in the network varies with the failed link. As a result, different allocations get activated in the network for different link failures. Due to this, FS `3' of lightpath ${l}_{1}$ at XC `2' experiences different interference power for different link failures. This scenario is same for each FS in every link at all XCs in the network. Now, to have a QoT guaranteed allocation, FS `3' can only be allocated to any request if and only if end-to-end signal to interference plus noise ratio (SINR) at FS `3' is greater than the required threshold (corresponding to targeted BER of $ 10^{-9} $) of considered modulation format (MF). Since we do not know which link of the network will fail beforehand, we employ robust optimization based RSA wherein we calculate end to end interference at FS `3' by considering each link failure and obtain maximum value among them. As we are considering the worst case interference scenario, QoT is guaranteed irrespective of the failed link. We follow the similar procedure before allocating each FS in working and backup path of every request. Implementing this robust optimization based RSA through MILP and heuristic is a challenging task.
	
	Considering the above challenges, we list our major contributions in this study as given below:
	\begin{itemize}
		
		\item We formulate a robust MILP optimization model in presence of IXT along with ASE and beating noise terms for SBPP EON for benchmarking against single link failures.
		
		\item For realistic networks, we propose a novel robust optimization based heuristic for static and dynamic traffic.
		
		\item For the static heuristic, we propose novel probability of congestion based sorting technique which improves the shareability among backup FSs.
		
		\item In our optimization framework, we ensure minimum QoT guarantee in terms of BER for both working and backup paths under any single link failure and no link failure.
		
		\item We extend our MILP and heuristic to design QoT guaranteed RSA in SBPP EON against SRLG failures.
		
		\item Through simulations, we demonstrate that RSA design in SBPP EON without considering PLIs contributes 33.94-41.02 percentage of QoT failed requests at 70 Tbps load whereas our design results in ``zero" QoT failed requests without affecting blocking probability much.
		
		
		\item We employ the efficient spectrum allocation technique named bitloading \cite{behera2019impairment} in our RSA design which was not implemented for survivable EONs/SBPP EONs so far.
	\end{itemize}

	\section{MILP Formulation} \label{MILP}
	In this section, we provide the MILP design model for the problem discussed in previous section. In the formulation of MILP, we define our objective function and present the basic RSA constraints like spectrum continuity, contiguity and non overlapping constraints. In addition, we illustrate the backup spectrum sharing and other SBPP constraints in the design process. Further, we demonstrate how the minimum QoT is ensured in both working and backup paths under any single link failure and no link failure conditions through our MILP framework. Note that, we use the following assumptions through out our paper. 
	
	\vspace{0.1cm}
	\hspace{-0.6cm}\textbf{\textit{Assumptions:} }
	In this work, balanced heterodyne detection is considered for CO-OFDM system. In addition, we assume that primary and interfering lightpath signals have identical power spectral density. We also consider that dispersion is perfectly compensated. Further, the bandwidth of each FS is assumed to be 12.5 GHz where we choose the base data rate with BPSK as 10 Gbps in each FS with 25\% overhead. We also assume that each node consists of sufficient variable data rate modulators.
	
	Note that, MILP input parameters and output variables are represented as [i/p] and [o/p], respectively. We use the following indices for the rest of our paper. \\
	
	\vspace{-0.2cm}
	\hspace{-0.55cm}\textbf{Indices} \\
	$ r,r' \in R = \left\lbrace 1, 2,  \dots,  C \right\rbrace $ \hspace{0.77cm} Set of given requests \\
	$ l,e \in L=\left\lbrace 1, 2,  \dots,  E \right\rbrace$ \hspace{0.945cm} Set of links in the network \\  
	$ f = \left\lbrace 1, 2, \dots,  N \right\rbrace $ \hspace{1.79cm} Available FSs in each link\\
	$ m = \left\lbrace 1, 2, \dots,  M \right\rbrace $ \hspace{1.66cm} Available modulation formats\\

	\hspace{-0.6cm}\textbf{\textit{Objective:} }
	\vspace{-0.3cm}
	\begin{align}
		\label{obj}
		minimize  \hspace{0.5cm} \underset{l}{\sum}\sum_{f}
		f \times H_{f,l} 
	\end{align} 
	
	\vspace{-0.1cm} 
	
	The objective of our MILP is to minimize the sum of highest indexed FS used in each link of the network to minimize the fragmentation in all links. $H_{f,l}$ in \eqref{obj} is set to 1 if FS \textit{`f'} is the highest indexed allocated FS in link \textit{`l'}, else 0.   
	
	\vspace{0.25cm}
	
	\hspace{-0.5cm}\textbf{\textit{Constraints:} } \\
	
	\vspace{-0.25cm}
	Highest indexed allocated FS in each link \textit{`l'} can be calculated by using the constraints (\ref{constraint1}-\ref{constraint3}). The parameters used in these constraints are described below \\
	\boldsymbol{$H_{f,l}$} =1 if FS \textit{`f'} is the highest indexed allocated FS in link \textit{`l'}, else 0. [o/p] \\
	\boldsymbol{$X_{f,l}$} =1 if FS \textit{`f'} is allocated in link \textit{`l'}, else 0. [o/p]
	
	\vspace{-0.3cm}
	\begin{equation}
		\begin{aligned}
			\label{constraint1}
			H_{f',l}  \leq 1- \frac{\sum_{f=f'+1}^{N}X_{f,l}} {N}; \,
			\forall l,
			\forall f' =\left\lbrace 1, \dots, N-1 \right\rbrace 
		\end{aligned}
	\end{equation} 
	
	\begin{equation}
		\label{constraint2}
		\begin{aligned}
			H_{f',l} & \leq  X_{f',l} \, ; \;\;
			\forall l,  \forall f'=\left\lbrace 1, \dots, N \right\rbrace \\
		\end{aligned}
	\end{equation} 
	
	\vspace{-0.2cm}
	\begin{equation}
		\label{constraint3}
		\begin{aligned}
			\sum_{f'=1}^{N} H_{f',l} & \geq \frac {\sum_{f'=1}^{N} X_{f',l}} {N} \, ; \;\;
			\forall l   
		\end{aligned}
	\end{equation}

	\subsubsection{Working and backup path selection} For each request, we compute $K$  shortest working paths and for each working path, we find $K_{b}$ shortest backup paths (link disjoint to corresponding working path). The parameters required for these constraints are given below.\\
	
	\vspace{-0.35cm}
	
	\hspace{-0.5cm}\boldsymbol{$P_{r}^{w}:$} Set of working paths of request \textit{`r'}. [i/p] \\
	\boldsymbol{$P_{r,p}^{b}:$} Set of backup paths of working path $p \in P_{r}^{w}$. [i/p] \\
	\boldsymbol{$W_{p}^{r}$} = 1 if request \textit{`r'} selects working path \textit{`p'}, else 0. [o/p]\\ 
	\boldsymbol{$B_{p_{b},p}^{r}$} = 1 if request \textit{`r'} chooses backup path `$p_{b}$' of working path \textit{`p'}, else 0. [o/p]\\
	
	\vspace{-0.45cm}
	
	\begin{equation}
		\label{constraint4}
		\begin{aligned}
			\sum_{p \in P_{r}^{w}} W_{p}^{r} = 1 \, ; \;\; 
			\forall r
		\end{aligned}
	\end{equation}  
	
	\begin{equation}
		\label{constraint5}
		\begin{aligned}
			\sum_{p_{b} \in P_{r,p}^{b}} B_{p_{b},p}^{r} = W_{p}^{r}\, ; \;\; 
			\forall p\in P_{r}^{w}\, , \forall r 
		\end{aligned}
	\end{equation}  
	
	Constraint \eqref{constraint4} guarantees that each request \textit{`r'} selects one working path \textit{`p'} from set of its working paths. Next, constraint \eqref{constraint5} ensures that each request \textit{`r'} chooses one backup path `$p_{b}$' from set of backup paths of selected working path \textit{`p'}.

	\subsubsection{Spectrum and Modulation assignment} 
	\hspace{-0.5cm}The parameters associated with these constraints are defined as follows: \\
	
	\vspace{-0.35cm}
	
	\hspace{-0.5cm}\boldsymbol{$\rho^{r}:$} Integer, data rate demand of request  \textit{`r'}. [i/p] \\
	\boldsymbol{$ \gamma :$} Base data rate with BPSK (10 Gbps in our case). [i/p] \\	
	\boldsymbol{$W_{m,f,p}^{r}$} = 1 if FS \textit{`f'} with MF \textit{`m'} is allocated in working path \textit{`p'} of request \textit{`r'}, else 0. [o/p]\\		
	\boldsymbol{$B_{m,f,p_{b},p}^{r}$} = 1 if FS \textit{`f'} with MF \textit{`m'} is allocated in backup path `$p_{b}$' of working path \textit{`p'} of request \textit{`r'}, else 0. [o/p]	
	
	\vspace{-0.2cm}
	
	\begin{align}
		\label{constraint6}
		\textbf{$W_{p}^{r}$} \times \rho^{r}= \sum_{m}\sum_{f} \textbf{$m.\gamma.W_{m,f,p}^{r}$} \, ; \;\;	
		\forall p\in P_{r} ^{w} \, , \forall r 
	\end{align}
	\vspace{-0.5cm}
	\begin{align}
		\hspace{-0.3cm}
		\label{constraint7}
		B_{p_{b},p}^{r} \times \rho^{r}= \sum_{m}\sum_{f} \textbf{$m.\gamma.B_{m,f,p_{b},p}^{r}$} \, ; \:\: 	
		\begin{cases}
			\forall p_{b} \in P_{r,p}^{b} \, , \\ \forall p\in P_{r} ^{w} \, ,\forall r \\
		\end{cases} 
	\end{align}
	
	\vspace*{-0.05cm}
	
	Constraint \eqref{constraint6} selects the FSs to be allocated in selected working path \textit{`p'} to satisfy the data rate demand of request \textit{`r'} and it also decides the MF of each selected FS. Similarly, constraint \eqref{constraint7} determines the FSs and corresponding MFs to be assigned in chosen backup path `$p_{b}$' of selected working path \textit{`p'} for request \textit{`r'}. Note that, since we use the bitloading technique in spectrum allocation \cite{behera2019impairment}, different FSs can be allocated with different MFs in working and backup path of each request.
	
	\vspace{-0.4cm}
	
	\begin{align}
		\label{constraint8}
		\sum_{m} \sum_{p \in P_{r}^{w}} {W_{m,f,p}^{r}} \leq 1\, ; \;
		\forall f,r
	\end{align}
	\vspace{-0.2cm}
	\begin{align}
		\label{constraint9} 
		\sum_{m} \sum_{p \in P_{r}^{w}} \sum_{p_{b} \in P_{r,p}^{b}} {B_{m,f,p_{b},p}^{r}} \leq 1\,; \;
		\forall f,r 
	\end{align}
	
	Constraint \eqref{constraint8} ensures that each selected FS is assigned with only one MF in one chosen working path \textit{`p'} for each request \textit{`r'}. Likewise, constraint \eqref{constraint9} guarantees that only one MF is allocated for every chosen FS in one selected backup path `$p_{b}$' of chosen working path \textit{`p'} for each request \textit{`r'}.

	\subsubsection{Spectrum contiguity constraint}
	 
	\vspace{-0.7cm}
	\begin{align}
		\label{constraint10}
		W_{m,f+1,p}^{r} & \geq (W_{m,f+2,p}^{r}+W_{m,f,p}^{r}-1)/N \; ;\nonumber \\
		W_{m,f+1,p}^{r} & \leq {W_{m,f+2,p}^{r}}+{W_{m,f,p}^{r}} \; ; \nonumber \\&
		\hspace{1.9cm} 
		\begin{cases}
			\forall  p\in P_{r}^{w} \, , \forall m,f,r 
		\end{cases} 
	\end{align} 
	
	\vspace{-0.55cm}
	
	\vspace{-0.2cm}
	\begin{align}
		\hspace{0.55cm}
		\label{constraint11}
		B_{m,f+1,p_{b},p}^{r} & \geq (B_{m,f+2,p_{b},p}^{r}+B_{m,f,p_{b},p}^{r}-1)/N \; ;  \nonumber \\
		{B_{m,f+1,p_{b},p}^{r}} &\leq {B_{m,f+2,p_{b},p}^{r}}+{B_{m,f,p_{b},p}^{r}} \; ;  \nonumber  \\&
		\hspace{0.4cm} 
		\begin{cases}
			\forall p_{b} \in P_{r,p}^{b}  \, ,\forall  p\in P_{r}^{w} \, , \forall m,f,r 
		\end{cases} 
	\end{align}
	
	Constraints \eqref{constraint10} and \eqref{constraint11} ensure the contiguity constraint of EON for working and backup path of each request \textit{`r'}. If FS \textit{`f'} is chosen in path, then constraints \eqref{constraint10} and \eqref{constraint11} make sure that FSs $ f+1,f+2 $ and so on will be selected till the requested data rate is satisfied.

	\subsubsection{Spectrum non-overlapping working and backup paths} \label{spectrum non-overlapping}
	\hspace{-0.55cm}The parameters employed in these constraints are described in the following. \\
	\hspace{-0.3cm}\boldsymbol{$\delta_{l, p}^{r}$} = 1 if working path \textit{`p'} of request \textit{`r'} contains the link \textit{`l'}, else 0. [i/p]\\
	\boldsymbol{$\delta_{l,p_{b},p}^{r}$} = 1 if backup path `$p_{b}$' of working path \textit{`p'} of request \textit{`r'} contains the link \textit{`l'}, else 0. [i/p] \\
	\boldsymbol{$W_{m,f,l}^{r}$} = 1 if FS \textit{`f'} with MF \textit{`m'} is allocated
	in link \textit{`l'} of the working path of request \textit{`r'}, else 0. [o/p]\\
	\boldsymbol{$B_{m,f,l}^{r}$} = 1 if FS \textit{`f'} with MF \textit{`m'} is allocated
	in link \textit{`l'} of the backup path of request \textit{`r'}, else 0. [o/p]\\
	\boldsymbol{$B_{f,l}$}=1 if backup path of at least one request $ r \in R $ is allocated with FS \textit{`f'} in link \textit{`l'}, else 0. [o/p]\\
	\boldsymbol{$X_{f,l}$}=1 if FS \textit{`f'} is allocated in link \textit{`l'}, else 0. [o/p]\\
	
	\vspace{-0.75cm}
	\begin{align}
		\label{constraint13}
		\hspace{-0.5cm}
		\sum_{p \in P_{r}^{w}} W_{m,f,p}^{r}  \times \delta_{l,p}^{r}=W_{m,f,l}^{r}\, ;  \;\;
		\forall \, m, f, l, r
	\end{align}
	
	\vspace*{-0.4cm}
	\begin{align}
		\label{constraint14}
		\sum_{p \in P_{r}^{w}} \sum_{p_{b} \in P_{r,p}^{b}} \textbf{$B_{m,f,p_{b},p}^{r}$}  \times \delta_{l,p_{b},p}^{r}=B_{m,f,l}^{r} \, ; 
		\forall \, m,f,l,r
	\end{align}
	
	\vspace{-0.3cm}
	
	\begin{align}
		\label{constraint15}
		\frac{\sum_{r}\sum_{m} B_{m,f,l}^{r}}{N} 
		\leq B_{f,l} \, ; \forall \, f,l
	\end{align}
	
	\vspace{-0.4cm}
	\begin{align}
		\label{constraint16}
		\sum_{r} \sum_{m} W_{m,f,l}^{r} \, +  B_{f,l} \, \leq \,X_{f,l}\, ;   
		\forall \, f,l
	\end{align}

	Constraints \eqref{constraint15} and \eqref{constraint16} enforce that working path of one request \textit{r} and backup path of other request $r'$ do not use the same FS(s) in their common links.  
	
	\subsubsection{Spectrum non-overlapping working paths} 
	Eq. \eqref{constraint16} also ensures that working paths of two different requests do not use the same FS(s) in their common links.
	
	\subsubsection{Backup sharing \& Link disjoint constraint} \label{link_disjoint_constraint}
	Here in this MILP framework, we do not include the constraint for spectrum non-overlapping between backup paths of any two requests. This enables the sharing of spectrum resources among the backup paths. Further, since our objective function tries to do the allocation at left side of the spectrum, backup paths are allocated with maximum possible spectrum sharing. However, backup paths of requests can not share the spectrum unless their corresponding working paths are link disjoint. Therefore, we provide the link disjoint constraints in (\ref{constraint17}) and (\ref{constraint18}). Parameter $t^{r,r'}$ in (\ref{constraint17}) gives the value 1 if two different requests $r$ and $r'$ has at least one common link in their working paths, else 0. Then Eq. (\ref{constraint18}) ensures that spectrum sharing among the backup paths is possible only when their corresponding working paths are link disjoint.
	
	\vspace{-0.3cm}
	\begin{align}
		\label{constraint17}
		t^{r,r'}\, & \geq \, \frac{\overset{M}{\underset{m,m'=1}{\sum}} \: 
			\overset{N}{\underset{f,f' =1}{\sum}} \: \underset{l}{\sum} \,\, W_{m,f,l}^{r}\times W_{m',f',l}^{r'}}{N} \, ; \, \nonumber \\ 
		t^{r,r'}\, &\leq \, \overset{M}{\underset{m,m'=1}{\sum}} \: \overset{N}{\underset{f,f' =1}{\sum}} \: \underset{l}{\sum} \,\, W_{m,f,l} ^{r} \times W_{m',f',l} ^{r'} \, ; \, \nonumber \\ &
		\hspace{3.7cm}
		\begin{cases} 
			\forall r,r'\, \text{where}, r \neq r' 
		\end{cases}
	\end{align}
	
	\vspace{-0.5cm}
	\begin{align}
		\label{constraint18}
		\hspace{-1.4cm}
		\sum_{m=1}^{M} B_{m,f,l}^{r} \, + \, \sum_{m'=1}^{M} B_{m',f,l}^{r'} \, \leq \, (2-t^{r,r'}).X_{f,l}\, ; \nonumber 
		\\
		\forall \, f,l,r,r' \, \text{where}, r \neq r'  
	\end{align}
	
	\subsubsection{Spectrum continuity constraint} Constraints \eqref{constraint6}, \eqref{constraint7}, \eqref{constraint8}, \eqref{constraint9}, \eqref{constraint13}, \eqref{constraint14}, \eqref{constraint15} and \eqref{constraint16} also ensure that  same FSs are allocated in all links of each working and backup path.
	\vspace{0.02cm}
	\subsubsection{Single link failure constraint} \label{link_failure} Constraint \eqref{constraint19} fails a single link \textit{`e'} in the network if there is a failure. Further, constraint \eqref{constraint20} denotes the  failed requests due to the failure of link \textit{`e'}. The following parameters are employed for these constraints. \\
	\hspace{-0.4cm}\boldsymbol{$F_{e}$} = 1 if link \textit{`e'} is failed in the network, else 0. [o/p]\\
	\boldsymbol{$F_{e,p}^{r}$} = 1 if failed link \textit{`e'} is present in working path \textit{`p'} of request \textit{`r'}, else 0. [o/p]\\
	\boldsymbol{ $\delta_{e, p}^{r}$} = 1 if working path \textit{`p'} of request \textit{`r'} contains the link \textit{`e'}, else 0. [i/p]\\
	\vspace{-0.5cm}
	\begin{align}
		\label{constraint19}
		\hspace{-0.7cm}
		\sum_{e \in L} F_{e}=1 \, ; \;
	\end{align}
	\vspace{-0.5cm}
	\begin{align}
		\label{constraint20}
		\hspace{0.1cm}
		F_{e} \times \delta_{e,p}^{r}=F_{e,p}^{r} \, ; \; 
		\forall p \in P_{r}^{w} \, ,  \forall e \in L, \forall r 
	\end{align}
	
	As aforementioned, each FS in working and backup path of every request experiences different IXT interference effect for different single link failures and no link failure conditions due to the presence of different active working and backup paths. Since we do not know which link of the network will fail beforehand, robust optimization based RSA is required to ensure the QoT under any link failure scenario. For this purpose, we fail single link at a time in the network using constraint (\ref{constraint19}) and calculate the end-to-end interference at FSs in working and backup path (whichever is active) of every request due to the active working and backup paths of other requests correspond to failed link. We repeat the above procedure for each link failure case and no link failure case and calculate the maximum possible (worst case) end-to-end interference among the above mentioned scenarios at FSs in working and backup path of every request. Now, the allocation is done based on worst case interference value at each FS of existing requests and incoming requests. Since we are considering worst case interference, whichever link fails, QoT is guaranteed in working and backup paths of each request. Above procedure is implemented using following constraints.
	
	\vspace{0.05cm}
	\subsubsection{Maximum possible crosstalk interference calculation in working paths} 
	\label{interference_calculation_working} 
	
	Crosstalk interference power generated at FS \textit{`f'} with MF \textit{`m'} in link \textit{`l'} of targeted working path \textit{`p'} of request\textit{ `r'} for the failure of link \textit{`e'} ($ l \neq e $) is equal to the sum of all crosstalk powers due to FS \textit{`f'} used in active working and backup paths of other requests which are
	passing through the head node of link \textit{`l'} (denoted by\textit{`i'}) as shown in \eqref{constraint21}. This computation is done for all FSs in each link of all working paths of all requests by considering each link failure \textit{`e'} in the network except the links present in targeted working path.
	
	\vspace{-0.5cm}
	\begin{align}
		\hspace{-1.2cm}
		\text{Pxt}_{m,f,p,l,e}^{w,r}=  
		\sum_{r' \in R/r}\sum_{ p' \in P_{r'}^{w}}\sum_{m'=1}^{M}\sum_{j'=1}^{N} \, \bigg[ {P}_{m',f,p',j',i}^{r'}.W_{m',f,p'}^{r'}.(1-F_{e,p'}^{r'})+ \nonumber 
	\end{align}
   \vspace{-0.5cm}
	\begin{align}
	\hspace{5cm}
	\label{constraint21}
	\sum_{p'_{b} \in P_{r',p'}^{b}} P_{m',f,p'_{b},p',j',i}^{r'}.B_{m',f,p'_{b},p'}^{r'}.F_{e,p'}^{r'}\bigg ] \, ;  \nonumber \\
	\forall e \in L/l_p,\forall l \in l_{p}, \forall p \in P_{r}^{w}, \forall \, m,f,r 
	\end{align}

	Here, $ l_p $ denotes the set of links present in working path \textit{`p'} and $P_{m',f,p',j',i}^{r'}$ is the power of interfering signal added at node \textit{`i'} (head node of link \textit{`l'}) which is using FS \textit{`f'} with MF $m^\prime$ and traverses from the node $j^\prime$ to the XC present in node \textit{`i'}, where $ j^\prime $-$ i $ is the link present in working path $p^\prime$ of request $r^\prime$. Next, the definition of $ P_{m',f,p'_{b},p',j',i}^{r'}$ in Eq. \eqref{constraint21} is same as $P_{m',f,p',j',i}^{r'}$ except that  $ j^\prime $-$ i $ is the link present in backup path $p'_{b}$ of working path $p^\prime$ of request $r^\prime$. $P_{m',f,p',j',i}^{r'}$ and $ P_{m',f,p'_{b},p',j',i}^{r'}$ are calculated as given below.
	 
	\vspace{-0.1cm}
	\begin{equation} 
		P_{m',f,p',j',i}^{r'}=P_{m',f,p'_{b},p',j',i}^{r'}=P_r \times C_x
	\end{equation}
	
	Where, $ P_r $ denotes the received signal power (we assume equal power for each FS) and $ C_x $ represents the crosstalk factor of the switch. Note that, $ P_{m',f,p',j',i}^{r'}=0 $ if no link exists between the nodes $j^\prime$ and \textit{i} in working path $p^\prime$ and
	$ P_{m',f,p'_{b},p',j',i}^{r'}=0$ if there is no link between $j^\prime$ and \textit{i} in backup path $p'_{b}$ of working path $p^\prime$.
	
	\vspace{0.1cm}
	Further, total aggregated crosstalk interference power at FS \textit{`f'} with MF \textit{`m'} at the destination node of targeted working path \textit{`p'} of request \textit{`r'} for the failure of link \textit{`e'} is computed by adding the crosstalk power at FS \textit{`f'} in each link of the path \textit{`p'} for the failure of link \textit{`e'} as shown in \eqref{constraint22}. The above calculation is repeated for the failure of each link \textit{`e'} in the network except the links present in targeted path \textit{`p'}. Next, the whole process is repeated for all FSs in each working path of every request.
	
	\vspace{-0.7cm}
	\begin{align}
		\hspace{-0.2cm}
		\label{constraint22}
		\text{Pxt}_{m,f,p,e}^{w,r}= \sum_{l \in l_{p}} \text{Pxt}_{m,f,p,l,e}^{w,r} \, ; 
		\begin{cases}
			\forall \,  e \in L/l_p, \forall p \in P_{r}^{w} \, , \\
			\forall \, m,f,r
		\end{cases}
	\end{align}
	
	Note that, $\text{Pxt}_{m,f,p,e}^{w,r}$ in constraint \eqref{constraint22} is different for different link failures due to the existence of different active working and backup paths. Therefore, maximum among all crosstalk power values that $\text{Pxt}_{m,f,p,e}^{w,r}$ constitutes considering each link failure \textit{`e'} is calculated as given in \eqref{constraint23}. This process is repeated for all FSs in each working path of every request.
	
	\vspace{-0.5cm}
	\begin{align}
		\label{constraint23}
		\hspace{-0.25cm}
		\text{Pxt}_{m,f,p}^{w,r,max_e}= \max_{e \in L/l_p}
		\left \{\text{Pxt}_{m,f,p,e}^{w,r} \right \}  ;
		\forall p \in P_{r}^{w} \, ,\forall \, m,f,r 
	\end{align}
	
	Further, we also enumerate the crosstalk interference power at all FSs of each working path of all requests under no link failure using the constraints (\ref{constraint24}-\ref{constraint25}).
	
	\vspace{-0.5cm}
	\begin{align}
		\label{constraint24}
		\text{Pxt}_{m,f,p,l}^{w,r}=\sum_{r' \in R/r}\sum_{ p' \in P_{r'}^{w}}\sum_{m'=1}^{M}\sum_{j'=1}^{N} \, \bigg[P_{m',f,p',j',i}^{r'}.W_{m',f,p'}^{r'} \bigg] \, ; \nonumber \\
	\forall l \in l_{p}, \forall p \in P_{r}^{w} \, , \forall \, m,f,r
	\end{align}

	\vspace{-0.4cm} 
	
	\begin{equation}
		\hspace{-1.7cm}
		\label{constraint25}
		\begin{aligned}
			\text{Pxt}_{m,f,p}^{w,r}= \sum_{l \in l_{p}}\text{Pxt}_{m,f,p,l}^{w,r} \, ; \;
			\forall p \in P_{r}^{w} \, , 
			\forall \, m,f,r 
		\end{aligned}
	\end{equation}

	Finally, maximum possible crosstalk power at each FS \textit{`f'} with MF \textit{`m'} at the destination node of every working path \textit{`p'} of each request \textit{`r'} considering all possible single link failure and no link failure conditions is computed as shown in \eqref{constraint26}.
	\vspace{-0.7cm}
	
	\begin{align}
		\label{constraint26}
		\text{Pxt}_{m,f,p}^{w,r,max}= 
		\max \left \{ \text{Pxt}_{m,f,p}^{w,r,max_e},\text{Pxt}_{m,f,p}^{w,r} \right \} \, ; 
			\forall p \in P_{r}^{w} \, ,\forall m,f,r
	\end{align}
	
	\subsubsection{Maximum possible crosstalk interference calculation in backup paths}
	\label{interference_calculation_backup} 
	By following the similar approach used for working path, we calculate the maximum possible crosstalk interference power at each FS in every backup path of all requests using the constraints (\ref{constraint27}-\ref{constraint29}). Note that, unlike for working paths, while finding maximum possible interference for backup paths, we fail one link of its working path as backup path is enabled when its working path fails. Next, the definitions of parameters used in these constraints are same as that we used for working paths in previous subsection. By replacing \textit{`w'} with \textit{`b'} and `$p$' (working path) with $ p_b,p $ (backup path `$p_b$' of working path `$p$') in the parameters used in previous section, we get the required description for the parameters used in these constraints.
	
	\vspace{-0.4cm}
	\begin{align}
		\hspace{-1cm}
		\text{Pxt}_{m,f,p_{b},p,l,e}^{b,r}=  
		\sum_{r' \in R/r}\sum_{ p' \in P_{r'}^{w}}\sum_{m'=1}^{M}\sum_{j'=1}^{N} \, \bigg[ {P}_{m',f,p',j',i}^{r'}.W_{m',f,p'}^{r'}.(1-F_{e,p'}^{r'})+ \nonumber 
	\end{align}
	\vspace{-0.5cm}
	\begin{align}
	\hspace{5.02cm}
	\label{constraint27}
	\sum_{p'_{b} \in P_{r',p'}^{b}}P_{m',f,p'_{b},p',j',i}^{r'}.B_{m',f,p'_{b},p'}^{r'}.F_{e,p'}^{r'}\bigg ] \, ; \nonumber \\  
	\forall e \in l_{p},
	\forall l \in l_{p_b,p},
	\forall p_{b} \in P_{r,p}^{b} \, ,  \forall p \in P_{r}^{w} \, ,
	\forall \, m,f,r  
	\end{align}

	\vspace{-0.4cm}
	
	\begin{align}
		\text{Pxt}_{m,f,p_{b},p,e}^{b,r}= \sum_{ l \in l_{p_b,p}} \text{Pxt}_{m,f,p_{b},p,l,e}^{b,r} \, ; 
		\forall e \in l_{p},\
		\forall p_{b} \in P_{r,p}^{b} \, ,  \forall p \in P_{r}^{w} \, ,  \forall \, m,f,r 
	\end{align} 
	
	\vspace{-0.5cm}
	
	\begin{align}
		\label{constraint29}
		\hspace{-1.1cm}
		\text{Pxt}_{m,f,p_{b},p}^{b,r,max}= \max_{e \in l_{p}} \left \{\text{Pxt}_{m,f,p_{b},p,e}^{b,r} \right \} \, ; 
			\forall p_{b} \in P_{r,p}^{b} \, ,  \forall p \in P_{r}^{w} \, ,
			\forall \, m,f,r 
	\end{align}
	
	\subsubsection{QoT guarantee constraints}
	\hspace{-0.55cm}The necessary parameters for these constraints are introduced in the following. \\
	\boldsymbol{$P_{ch}:$} Coherently received power at any FS. [i/p]\\
	\boldsymbol{$SNR_{f,p}^{r}:$} Signal to noise ratio (SNR) at FS \textit{`f'} in working path \textit{`p'} of request \textit{`r'}. [i/p]\\
	\boldsymbol{$\sigma_{lo-sp}^{2\lvert(r,p)}:$} Local oscillator (LO)-ASE beat noise variance in working path \textit{`p'} of request \textit{`r'}. [i/p]\\
	\boldsymbol{$\zeta:$} $ (R_a^2/2) P_{lo}2\eta_{sp}hf_cB_e $ (refer Table \ref{sim_par} for each term). [i/p] \\
	\boldsymbol{$ M_p^{r}:$} Number of EDFAs required in working path \textit{`p'} of request \textit{`r'}. [i/p] \\
	\boldsymbol{$ E_s: $} EDFA spacing. [i/p]\\ 
	\boldsymbol{$ \sigma_{lo-x}^{2\lvert(w,r,m,f,p,max)}:$} Maximum possible 
	local oscillator (LO)-crosstalk beat noise variance at FS \textit{`f'} using MF \textit{`m'} in working path \textit{`p'} of request \textit{`r'}. [o/p]\\
	\boldsymbol{$IST_m^w:$} Denotes value of inverse of signal-to-interference-plus-noise-ratio (SINR) threshold for selected working path for MF \textit{`m'} (refer \cite{behera2019impairment} for BER expression). [i/p] \\
	\boldsymbol{$LN:$} Large number $ >>IST_M^w $ ; where, $ M $ is the maximum available MF. [i/p] \\ 
	
	\vspace{-0.3cm}
	Parameters related to working path \textit{`p'} are described above. By replacing \textit{`w'} with \textit{`b'} and `$p$' (working path) with $ p_b,p $ (backup path `$p_b$' of working path `$p$') in above parameters and corresponding description, we get the required parameters and description for the backup paths as well.
	
	To ensure minimum QoT guaranteed allocation, any FS is allocated in either working or backup path if and only if end-to-end SINR at that FS is greater than or equal to SINR threshold ($ SINR_{th} $) of considered MF. Therefore,

	\vspace{-0.4cm}
	\begin{align}
		\label{SINR_eq}
		SINR \geq SINR_{th} \;
		\implies P_{ch}/(P_I+P_N) \geq SINR_{th} \;\;\;\;\;\;\;\;\; \nonumber
	\end{align}
	\vspace{-0.4cm}
	\begin{align}
		\hspace{2.7cm} \implies \frac{P_I}{P_{ch}}+\frac{1}{SNR} \leq \frac {1}{SINR_{th}} \, ;
	\end{align}

	\vspace{0.1cm}
	\noindent where, $Pch$: Coherently received signal power; $P_I$: Interference power; $P_N$: Noise power. Eq. \eqref{SINR_eq} is the underlying equation for QoT guarantee constraints of (\ref{constraint32}) and (\ref{constraint33}).  
	
	The combined noise term includes LO-crosstalk ($ P_I $) and LO-ASE ($ P_N $) beat noises when the system is assumed to be operated well above the shot noise limit with high LO power (refer \cite{behera2018effect} for a complete analysis). Next, Each term in \eqref{SINR_eq} is calculated as follows:
	  
	\vspace{-0.3cm} 
	\begin{equation}
		\hspace{-2.9cm} 
		\label{SNR}
		\begin{aligned}
			&  \hspace{2.95cm} SNR_{f,p}^{r}=\frac{P_{ch}}{\sigma_{lo-sp}^{2\lvert(r,p)}}, \text{where}, { P_{ch}}=({R_{a}^2}/{2})P_{lo}P_{r},  \\
		\end{aligned}
	\end{equation}  
	 
	\vspace{-0.3cm}   
	\begin{equation}
		\hspace{2.2cm} 
		\label{variance}
		\begin{aligned}
			&\hspace{-2cm} \sigma_{lo-sp}^{2\lvert(r,p)}=\zeta \left(M_p^{r}(G_{in}-1) + \sum_{i=1}^{Z-1} (G_{out}(i)-1)\right)
		\end{aligned} 
	\end{equation}
	
	\hspace{2.4cm}  $ \text{where}, M_p^{r}=\frac{\sum_{l \in l_p}\triangle_{l}}{E_s} \, ; \triangle_{l}: \text{link distance} $ \\
	
	\vspace{-0.15cm}
	In Eq. \eqref{variance}, $ G_{in} $ denotes the input EDFA gain and $G_{out}(i)$ represents the output EDFA gain at node $i=1,2,...,Z$ in working path `$p$' of request \textit{`r'} and given as \cite{behera2019impairment}, 
	$G_{out}(i) \geq 3 \ceil{log_{2}Q(i)}+L_{WSS} \hspace{0.2cm} \text{dB}$,  where $Q(i)$ is the fiber inputs/outputs at node `$i$'. Each parameter in Eq. \eqref{variance} is related to broadcast and select node architecture presented in \cite{behera2019impairment} which is employed for this work as well. 
	
	Similarly, $ SNR_{f,p_b,p}^{r} $ and $ \sigma_{lo-sp}^{2\lvert(r,p_b,p)} $ are computed by replacing `$p$' (working path) with $ p_b,p $ (backup path `$p_b$' of working path `$p$') in \eqref{SNR} and \eqref{variance}, respectively. Further, LO-crosstalk beat noise terms for working and backup paths are enumerated as given in \eqref{Lo-crosstalk1} and  \eqref{Lo-crosstalk2}, respectively.
	
	\vspace{-0.3cm}
	\begin{align}
	\hspace{1.5cm}
		\label{Lo-crosstalk1}
		& \hspace{-2.3cm}\sigma_{lo-x}^{2\lvert(w,r,m,f,p,max)} =({R_{a}^2}/{2}){P_{lo}
			{\mathrm{Pxt}_{m,f,p}^{w,r,max}}  
		} 
	\end{align}
	\vspace{-0.5cm}
	\begin{align}
	\hspace{1.5cm}
		\label{Lo-crosstalk2}
		& \hspace{-2cm}\sigma_{lo-x}^{2\lvert(b,r,m,f,p_b,p,max)} =({R_{a}^2}/{2}){P_{lo}
			{\mathrm{Pxt}_{m,f,p_b,p}^{b,r,max}}}  
	\end{align}
	
	\noindent where, $\mathrm{Pxt}_{m,f,p}^{w,r,max}$ and $ \mathrm{Pxt}_{m,f,p_b,p}^{b,r,max} $ 
	are from  \eqref{constraint26} and \eqref{constraint29}, respectively.
	
	Now, QoT guarantee constraints for working and backup paths are formulated as shown in \eqref{constraint32} and \eqref{constraint33}, respectively.
	
	\vspace{-0.2cm}
	\begin{align}
		\label{constraint32}
		\hspace{-1.5cm}
		\frac{\sigma_{lo-x}^{2\lvert(w,r,m,f,p,max)}}{P_{ch}} + \frac{W_{m,f,p}^{r}}{SNR_{f,p}^{r}} + LN\, \times \, W_{m,f,p}^{r} 	\leq LN+IST_{m}^{w}\, ; \; \nonumber \\ \forall p \in P_{r}^{w} \, ,
		\forall \, m,f,r 
	\end{align}
	
	\vspace{-0.6cm}
	
	\begin{align}    
		\label{constraint33}
		\frac{\sigma_{lo-x}^{2\lvert(b,r,m,f,p_b,p,max)}}{P_{ch}} + \frac{B_{m,f,p_{b},p}^{r}}{SNR_{f,p_b,p}^{r}} + LN \times {B_{m,f,p_{b},p}^{r}}  
		\leq LN + IST_{m}^{b} \, ; \; \nonumber \\
			\forall p_{b} \in P_{r,p}^{b} \, ,  \forall p \in P_{r}^{w} \, , 
			\forall \, m,f,r 
	\end{align}
	
	With respect to working path, when a particular $W_{m,f,p}^{r} = 1$, then Eq. \eqref{constraint32} is similar to \eqref{SINR_eq}. On the other hand, when $W_{m,f,p}^{r} = 0$, then $LN$ makes sure that Eq. \eqref{constraint32} is valid. Above description also holds for backup path QoT guarantee constraint given in \eqref{constraint33}. Thus, constraint \eqref{constraint32} (or \ref{constraint33}) ensures that, any FS is allocated in the working (or backup) path only when the worst case end-to-end SINR (considering each link failure scenario) at respective FS satisfies the SINR threshold criterion of the considered MF. Note that, worst case end-to-end SINR at any FS in working (or backup) path corresponds to maximum possible interference at that FS which is calculated in subsections \ref{interference_calculation_working} and \ref{interference_calculation_backup}, respectively. Since we considered worst case interference, minimum QoT assured allocation is guaranteed in both working and backup paths under any single link failure and no link failure conditions.
	
	It is worth mentioning that, Equations \eqref{constraint32} and \eqref{constraint33} which gets reflected to \eqref{constraint26} and \eqref{constraint29} via  \eqref{Lo-crosstalk1} and \eqref{Lo-crosstalk2} consists a maximization problem over interference powers corresponding to each link failure which is a random event. This makes our problem a \textit{robust optimization problem} which contains maximization in the constraints and minimization in the objective. Therefore, our problem is not a regular ILP to solve and has to be converted to normal ILP by changing maximization constraint into linear as shown in the appendix.
	
	Noteworthy, nonlinear constraints formulated in our MILP (Eq. \eqref{constraint17}, \eqref{constraint18}, \eqref{constraint21}, \eqref{constraint23}, \eqref{constraint26}, \eqref{constraint27}, \eqref{constraint29}) are converted into linear by following the standard procedure shown in appendix.
	
	\section{Heuristic Algorithm} \label{heuristic}
	It is proven that RSA without impairments in EON is NP-Hard problem \cite{behera2018effect}. The computational complexity of our proposed MILP increases exponentially with size and traffic of the network. In view of this, we propose a near optimal heuristic algorithm for realistic large network topologies. In this section, we propose a novel sorting technique to order the given static traffic requests followed by a heuristic algorithm to solve our problem under static and dynamic traffic scenarios. Note that, indices and parameters used in MILP are also employed in this section.
	
	\vspace{-0.2cm}
	\subsection{Sorting of requests for static traffic} \label{sorting_requests}
	It is worth mentioning that, sorting the requests is an integral part of any heuristic design for static traffic scenario as it influences the performance of heuristic in terms of optimality gap (G). \textit{`G'} is the percentage deviation between MILP objective value and that of heuristic. Therefore, sorting technique should be such that the value of \textit{`G'} is as minimum as possible. Note that, existing techniques sort the requests without taking the network congestion into account and there were no techniques so far which improves the shareability among backup FSs which results in high `G' value. Therefore, in this work we propose a novel sorting technique named most congested working-least congested backup first (MCW-LCBF). In this technique we calculate the probability of congestion in each working and backup path of every request from each working and backup path of other requests. Since more congestion is disadvantageous for working path allocation and advantageous for backup path allocation, we sort the requests in such a way that more sharing of FSs among backup paths takes place. We describe the proposed sorting technique with the help of following steps.
	
	
	\vspace{0.2cm}
	\subsubsection{ MCW-LCBF sorting technique }
	\hspace{-0.55cm}Required parameters for the proposed sorting technique are as follows:\\
	\hspace{-0.45cm} \boldsymbol{$l_{p}:$} Set of links present in working path `$p$'. \\
	\boldsymbol{$l_{p_b,p}:$} Set of links present in backup path `$p_b$' of working path `$p$'.\\
	\boldsymbol{$x^{l,p,r}_{w_{p'},r'}:$} Probability that the link \textit{`l'} of working path \textit{`p'} of request  \textit{`r'} gets congestion from working path  $p'$
	of request $r'$, where, $ r,r' \in R $ and $r \neq r'$. \\
	\boldsymbol{$x^{l,p,r}_{b_{p'},r'}:$} Probability that the link \textit{`l'} of working path \textit{`p'} of request  \textit{`r'} 
	gets congestion from backup paths of working path $p'$ of request $r'$, where,  $ r,r' \in R $ and $r \neq r'$. \\
	\boldsymbol{$ x_{r'}^{l,p,r}: $} Probability that the link \textit{`l'} of working path \textit{`p'} of request \textit{`r'} gets congestion from working and backup paths of request $r'$, where, $ r,r' \in R $ and $r \neq r'$. 
	
	\hspace{-0.5cm}\textbf{Step 1}: For each request ($r \in R$), we find $K$ shortest working paths and for each working path, we obtain $K_b$ shortest backup paths (link disjoint to corresponding working path). \\
	\textbf{Step 2:} Probability of congestion in each link present in every working path of each request ($ r \in R $) from all other requests ($ r' \in R/r$) is evaluated as follows: \\
	
	\vspace{-0.3cm}
	
	If link \textit{`l'} of working path \textit{`p'} of request \textit{`r'} goes through the working path $p'$ of request $r'$, then \textit{`l'} gets congestion from $p'$ if and only if request $r'$ selects $p'$ as its working path with probability $1/K$ as there are $K$ working paths for $r'$. Therefore,
	\vspace{-0.3cm}


	\begin{align}
	\label{probability_dueto_working}
	\hspace{-0.2cm}x^{l,p,r}_{w_{p'},r'}=
	\begin{cases}
	\frac{1}{K}  &\text{if link} \, \textit{`l'} \, \text{of working path \textit{`p'} of request} \, \textit{`r'} \\& \text{is} \, \, \text{present} \, \, \text{in working path} 
	\, \, \,
	p' \, \,  \text{of request} \, \, r' \\ & \text{where}, \, r \neq r' \, \text{and} \, \, p'\in P^w_{r'}  \\
	0 & \text{otherwise}  
	\end{cases}
	\end{align}
	
	Next, if link \textit{`l'} of working path \textit{`p'} of request \textit{`r'} goes through any one of the backup paths of working path $p'$ of request $r'$, then \textit{`l'} gets congestion from that backup path if $r'$ selects it as its backup path. For this, first, $r'$ has to select working path $p'$ with probability $1/K$ and then choose one among $K_b$ backup paths with probability $1/K_b$. Therefore, the total probability is    
	$(1/K)*(1/K_b) $. If \textit{`l'} goes through $ D_{p'}^{l} $ number of backup paths of $ p' $ of $r'$, probability will be $(1/K*1/K_b*D_{p'}^{l})$ as in (\ref{probability_dueto_backup}).
	\vspace{-0.1cm}
	\begin{align}
		\label{probability_dueto_backup}
		\hspace{-0.1cm}x^{l,p,r}_{b_{p'},r'}= & \frac{1}{K} *\frac{1}{K_b}*D_{p'}^{l} \, ; \,\text{where, $D_{p'}^{l}$ denotes the number of}\nonumber  \\ 
		& \text{backup paths (contains the link \textit{`l'} of working path} \nonumber\\&  \text{\textit{`p'} of request \textit{`r'}) of working path $p'$ of request $r'$,} \nonumber \\ & \text{where},
		r \neq r' \,
		\text{and} \, \, p'\in P^w_{r'} 
	\end{align}
	
	Sum of the probabilities in (\ref{probability_dueto_working}) and (\ref{probability_dueto_backup}) over all working paths $ p' $ of request $r'$ gives the total probability that link \textit{`l'} of working path \textit{`p'} of request \textit{`r'} gets congestion from $r'$ as given in \eqref{probability_calculation}.
	
	\vspace{-0.1cm}
	\begin{equation}
		\label{probability_calculation}
		x_{r'}^{l,p,r}=\sum_{p' \in P_{r'}^w} (x^{l,p,r}_{w_{p'},r'}+x^{l,p,r}_{b_{p'},r'}) \, ; 
	\end{equation}
	
	Now, the probability of congestion at any FS \textit{`f'} (CSL) in link \textit{`l'} of working path \textit{`p'} of request \textit{`r'} from all other requests $r' \in R/r$ is given as
	\vspace{-0.05cm}
	\begin{align}
		\label{congestion_slot_working}
		\hspace{-0.1cm}
		CSL^{l,p}_{r} = \underset{r' \in R/r}{\sum}\frac{x_{r'}^{l,p,r} \ast  \rho_{r'}^{BPSK}}{N} \, ; 
	\end{align}
	
	$\rho_{r'}^{BPSK}$ used in above equation represents the number of FSs required for requested data rate of each request ($\rho_{r'}$) when BPSK modulation is used for each FS. This means, we are considering the FSs allocation with uniform modulation using BPSK to consider the worst case scenario.
	
	Finally, probability of congestion in link \textit{`l'} of working path \textit{`p'} of request \textit{`r'} from all other requests $r' \in R/r$ is computed as
	\begin{align}
		\label{congestion_link_working}
		\hspace{-0.1cm}
		con^{l,p}_{r} = CSL^{l,p}_{r} \ast \rho_{r}^{BPSK} \, ; 
		\forall l \in l_{p}, 
		\forall p \in P^w_r,  \forall r
	\end{align}
	
	\hspace{-0.35cm}\textbf{Step 3:} By aggregating the congestion in corresponding links, probability of congestion in each working path of every requests is computed as shown below. \\
	\vspace{-0.2cm}
	\begin{equation}
		con^{p}_{r}= \underset{l \in l_{p}}{\sum} con^{l,p}_{r} \, ; \forall p \in P^w_r, \forall r  
		\label{congestion_working} 
	\end{equation}
	
	\hspace{-0.37cm}\textbf{Step 4:} By following similar approach in \eqref{congestion_slot_working} and \eqref{congestion_link_working}, we calculate the probability of congestion in each link present in every backup path of each working path of all requests.\\
	\vspace{-0.2cm}
	\begin{align}
		\label{congestion_slot_backup}
		CSL^{l,p_b,p}_{r} = \underset{r' \in R/r}{\sum} \frac{x_{r'}^{l,p_b,p,r} \ast  \rho_{r'}^{BPSK}}{N} \, ;
	\end{align}
	\vspace{-0.3cm}
	\begin{align}
		\label{congestion_link_backup}
		con^{l,p_b,p}_{r} = CSL^{l,p_b,p}_{r} \ast \rho_{r}^{BPSK} \, ;
		\begin{cases} 
			\forall l \in l_{p_b,p}, 
			\forall p_b \in P^b_{r,p}, \\ \forall p \in P^w_{r}, 
			\forall r 
		\end{cases}
	\end{align}
	
	$ x_{r'}^{l,p_b,p,r} $ is calculated using (\ref{probability_dueto_working}), (\ref{probability_dueto_backup}) and (\ref{probability_calculation}) by replacing `$p$' (working path) with $ p_b,p $ (backup path `$p_b$' of working path `$p$').
	
	\hspace{-0.35cm}\textbf{Step 5:} Probability of congestion in each backup path of every working path of all requests is evaluated.\\
	\begin{equation}
		\label{congestion_backup}
		con^{p_b,p}_{r}= \underset{l \in l_{p_b,p}}{\sum} con^{l,p_b,p}_{r} \, ;
		\forall p_b \in P^b_{r,p}, 
		\forall p \in P^w_{r}, \forall r
	\end{equation}
	
	\hspace{-0.3cm}\textbf{Step 6:}
	In general, from working path point of view, requests with more congestion in their working path should be allocated first as highly congested requests might not get enough resources if we don't allocate them first. Conversely, from backup path point of view, requests with less congestion in their backup path should be allocated first. This is because, since we ensure the sharing of spectrum in backup path, the lightly congested request has less chances of sharing the spectrum with other requests and thus difficult to get the resources if we allocate at the end. Therefore, to have a unified parameter for sorting, in this step, we calculate the ratios of individual working path's congestion to each of its corresponding backup path's congestion for every request.
	\begin{equation}
		\frac{con_{r}^{p}}{con_{r}^{p_b,p}} \, ;
		\forall p_b \in P^b_{r,p}, \forall p \in P^w_{r},
		\forall r \label{finding_ratios}
	\end{equation}
	
	\hspace{-0.3cm}\textbf{Step 7:} We compute the average value of corresponding ratios of each request to find out the average probability of congestion for every request.
	\vspace{-0.1cm}
	\begin{equation}
		\label{congestion_request}
		con_{r}= \frac{1}{\left(K \ast K_b\right)} \left[\sum_{ p \in P^w_{r}} \sum_{p_b \in P^b_{r,p}} \frac{con_{r}^{p}}{con_{r}^{p_b,p}}\right] \, ;
		\forall r 
	\end{equation}
	\textbf{Step 8:} Requests are sorted in descending order of their average probability of congestion evaluated in (\ref{congestion_request}).
	
	\begin{algorithm} [h!]
		\caption{SBPP-IA Heuristic Algorithm} \label{working_algorithm}
		\begin{algorithmic}[1]
			\State Sort all the requests according to MCW-LCBF 
			\For{each request \textit{`r'} } 
			\State Find $K$ shortest working paths and for each working path, calculate \hspace*{0.5cm} $K_b$ backup paths (link disjoint to corresponding working path) using  \hspace*{0.5cm} Dijkstra's algorithm
			\State $ n_s = \ceil[\bigg]{\frac{\rho_{r}}{10 \ast M}} ; \,
			\forall p \in P_{r}^{w}, \,
			\forall p_b \in P_{r,p}^{b} $ \label{MFS}
			\State Obtain $CFS^{p}_{r}$ and $CFS^{p_b,p}_{r}$ ; $\forall p \in P_{r}^{w}, \, 
			\forall p_b \in P_{r,p}^{b} $ \label{CFS_arrays}
			\For{each working path $p \in P_{r}^{w}$}
			\State $Y \leftarrow \text{Empty array}$, $i=1$
			\State $\rho'_{r}=\rho_{r}$ \label{rho}
			\State $f= CFS^{p}_{r}(i)$ \label{FS}
			\For{failure of each link $e \in L/l_p$} \label{int_link_start}
				\For{each link \textit{l} $\in l_{p}$} \label{path_allocation}
				\State Calculate IXT interference at \textit{`f'} using \eqref{constraint21} as, $int^{f,e}_{l}$ \label{int_link}
				\EndFor
				\State Calculate end-to-end IXT interference at \textit{`f'} as, $int^{f,e}_{p,r}=\underset{l  \in l_{p}}{\sum} int^{f,e}_{l}$ \label{int_path}
				\vspace{-0.6cm}
				\EndFor \label{int_link_end}
				\State Calculate maximum of all end-to-end IXT interferences at \textit{`f'} corre- \hspace*{1.2cm}sponding to each link failure as,  $int^{f,max_e}_{p,r}= \underset{e \in L/l_p}{max}\left(int^{f,e}_{p,r}\right)$ \label{max_int_path}
				\State Calculate end-to-end IXT interference at \textit{`f'} in no failure case \hspace*{1cm} using \eqref{constraint24} and \eqref{constraint25} as,  $int^{f,nf}_{p,r}$ \label{nlf_int_path}
				\State Calculate maximum possible IXT interference (considering each link \hspace*{1.2cm}failure scenario) at \textit{`f'} as, $int^{f,max}_{p,r}=\textit{max} \left (int^{f,max_e}_{p,r},int^{f,nlf}_{p,r}\right )$ \label{final_max_int_path}
				\State Calculate maximum possible IXT interference at \textit{`f'} in working ($ p'$) \hspace*{1cm} and backup path ($ p'_{b}$) of existing requests ($ R_{exi} $) considering the \hspace*{1cm} allocation of \textit{`f'} in \textit{`p'} as,  $int^{f,max}_{p',r'}$ \& $int^{f,max}_{p'_b,p',r'}$ $ \forall r' \in R_{exi} $  \label{int_existing}
				\State Calculate ASE noise in \textit{`p'} as $noi^{f}_{p,r}$ and in $p'$, $ p'_b $  of $ R_{exi} $ as $noi^{f}_{p',r'}$ \hspace*{1cm} \& $noi^{f}_{p'_b,p',r'}$ $ \forall r' \in R_{exi} $ \label{ASE_calculation} 
				\algstore{myalg}
		\end{algorithmic}	
	\end{algorithm} 
	
	\begin{algorithm} [h!]
		\begin{algorithmic}[1]
			\algrestore{myalg}
				\State Compute worst case end-to-end SINR (considering each link failure \hspace*{1cm} scenario) at \textit{`f'} in \textit{`p'} as  $SINR^{f,min}_{p,r}$  and at \textit{`f'} in  $p'$, $ p'_b $ of $ R_{exi}$ as \hspace*{1.2cm}$SINR^{f,min}_{p',r'}$  \&  $SINR^{f,min}_{p'_b,p',r'} \forall r' \in R_{exi}$ \label{SINR_calculation}
			\State ($ X,Y,i,f $)=  \text{QoT}$(SINR^{f,min}_{p,r},SINR^{f,min}_{p',r'},SINR^{f,min}_{p'_b,p',r'},$
		   \Statex	\hspace*{9.2cm}$m,f,i,p,\rho'_{r},Y) $
			\If{X=1} goto step \ref{rho}
			\ElsIf{X=2} goto step \ref{int_link_start}
			\Else $ AFS \leftarrow Y $ \label{AFS}
			\EndIf
			\For{each backup path $p_b \in P_{r,p}^{b}$} \label{backup_allocation} 
			\State $Y \leftarrow \text{Empty array}$
			\State $AFS_b$=BackupPath($\rho_{r},p,p_b,CFS^{p_b,p}_{r},Y$) \label{calling_sub_routine}
			\State Calculate $obj^{p_b}_{p,r}$ \label{finding_objective} by considering allocations in existing requests \hspace*{1.5cm} and \textit{AFS},$AFS_b$ of current request
			\EndFor
			\EndFor
			\State $obj(r)=\underset{p,p_b}{min} \left(obj^{p_b}_{p,r}\right)$ \label{min_objective_value}
			\State Choose \textit{`p'} and `$p_b$' combination for which $obj^{p_b}_{p,r}$ is minimum and allocate \hspace*{0.5cm} assigned FSs with MFs \label{min_objective_value2}
			\EndFor
		\end{algorithmic}	
		\end{algorithm}

	\subsection{Proposed SBPP-Impairment Aware (SBPP-IA) Algorithm} \label{heuristic_algorithm}
	Requests are allocated one by one in a sequence given by the proposed sorting technique in previous subsection. We execute the following steps to establish each request. Since we employ bitloading technique \cite{behera2019impairment} for spectrum allocation wherein different FSs can be allocated with different MFs, number of FSs required to satisfy the data rate demand ($\rho_{r}$) of given request \textit{`r'} is unknown beforehand.
	Therefore, first we calculate the minimum number of FSs required ($n_s$) to establish \textit{`r'} using the equation $n_s = \ceil[\bigg]{\frac{\rho_{r}}{10 \ast M}}$ (line \ref{MFS}, Algorithm \ref{working_algorithm}), where, $ M $ is the highest available MF. 
	Next, we obtain an array $CFS^{p}_{r}$, $\forall p \in P^w_r$, in which each element denotes the FS number from where $n_s$ free contiguous FSs are present in working path \textit{`p'} of request \textit{`r'} (line \ref{CFS_arrays}, Algorithm \ref{working_algorithm}). Similarly we obtain $CFS^{p_b,p}_{r}, \, \forall p_b \in P_{r,p}^{b}, \forall p \in P^w_r$ which gives the details about set of free $n_s$ contiguous FSs in backup path `$p_b$' of working path \textit{`p'} of request \textit{`r'} (line \ref{CFS_arrays}, Algorithm \ref{working_algorithm}). Note that, $CFS^{p_b,p}_{r}$ is obtained in such a way that sharing of FSs among backup paths occur subject to the constraints in subsection \ref{link_disjoint_constraint}. Further, elements in $CFS^{p}_{r}$ and $CFS^{p_b,p}_{r}$ are sorted in ascending order using First-Fit algorithm \cite{behera2018effect} to minimize the fragmentation in each link of all working and backup paths. We block the request \textit{`r'} if $n_s$ free contiguous FSs are not free in at least one working path and one of its corresponding backup paths.

	\subsubsection{Allocation for working path}
	We begin with first FS \textit{`f'} in first set of $n_s$ free contiguous FSs in considered working path \textit{`p'} (line \ref{FS}, Algorithm \ref{working_algorithm}). Now, we calculate the worst case (robust) end-to-end SINR (considering each link failure case) at targeted FS \textit{`f'} in path \textit{`p'} and we also compute the worst case end-to-end SINR at \textit{`f'} in working ($ p' $) and backup path ($ p'_{b} $) of existing requests ($ R_{exi} $) considering the allocation of targeted FS \textit{`f'} in path \textit{`p'} (lines \ref{int_link_start}-\ref{SINR_calculation},  Algorithm \ref{working_algorithm}). If worst case SINR at \textit{`f'} in  $ p' $, $ p'_{b} $ of $ R_{exi} $ satisfies SINR thresholds of assigned MFs, we tentatively allocate FS \textit{`f'} with the highest available MF \textit{`m'} for which worst case SINR at \textit{`f'} in path \textit{`p'} is greater than SINR threshold of \textit{`m'} (lines  \ref{SINR_constraint}-\ref{positive_rho_start},\ref{zero_rho} Algorithm \ref{working_algorithm}). 
	After the allocation of  \textit{`f'}, if data rate demand of  \textit{`r'} is not satisfied, we go for next contiguous FS and repeat the same procedure for this new FS if it is free (lines \ref{positive_rho_start},\ref{slot_available}, Subroutine 2). Conversely, if this new FS is not free, we discard the previous allocations and start with next set of $n_s$ contiguous FSs with fresh $\rho_{r}$ (line \ref{slot_not_available}, Subroutine 2) and repeat the above steps. If SINR at \textit{`f'} in  $ p' $, $ p'_{b} $ of $ R_{exi} $ fails to satisfy required thresholds of assigned MFs or if SINR at \textit{`f'} in path \textit{`p'} is less than the SINR thresholds of all available MFs, we discard the previous allocations and start with next set of $n_s$ contiguous FSs with fresh $\rho_{r}$ (lines \ref{slot_not_available2},\ref{slot_not_available3}, Subroutine 2). The whole process is repeated till the requested data rate is allocated in considered path \textit{`p'}. 
	
	Once the required datarate is allocated, we stop this process without checking any further FSs in path \textit{`p'} and store the allocated FSs in the array AFS (line \ref{AFS}, Algorithm \ref{working_algorithm}). Next, we focus on allocation of backup paths of considered working path \textit{`p'} (line \ref{backup_allocation}, Algorithm \ref{working_algorithm}). However, if the required FS allocation is not possible in path \textit{`p'} we assign an empty array to AFS and consider the next working path.

	\begin{algorithm} [h!]
		\floatname{algorithm}{} 
		\renewcommand{\thealgorithm}{} 
		\caption{Subroutine 2: \text{QoT}$(SINR^{f,min}_{p,r},SINR^{f,min}_{p',r'},SINR^{f,min}_{p'_b,p',r'},m,f,i,p,\rho'_{r},Y) $
			\label{QoT_check}}
		\begin{algorithmic}[1]
			\If{$SINR^{f,min}_{p',r'} \geq SINR^{m'}_{th}(f,p',r')$ \&
				$SINR^{f,min}_{p'_b,p',r'}\geq SINR^{m''}_{th}(f,p'_b,p',r') \,  \forall r' \text{ where,} \, SINR^{m'}_{th}(f,p',r') $ denotes SINR threshold of MF $ m' $ (to satisfy minimum QoT)  assigned for FS \textit{`f'} in working path $ p' $ of request $ r' $ and $ SINR^{m''}_{th}(f,p'_b,p',r') $ denotes SINR threshold of MF $ m'' $ assigned for FS \textit{`f'} in backup path $ p'_b $ of working path $ p' $ of request $ r' $} \label{SINR_constraint}
			\State Select highest available MF \textit{`m'} for which $SINR^{f,min}_{p,r} \geq \hspace*{0.45cm} SINR^{m}_{th}(f,p,r)$
			\If {any MF \textit{`m'} is available}
			\State $\rho'_{r}=\rho'_{r} - 10 \ast m$ \label{rho_after_allocation}
			\If{$\rho'_{r} < 0$} $\rho'_{r}=\rho'_{r} + 10 \ast m$, $m=m-1$, goto step \ref{rho_after_allocation} 
			\ElsIf{$\rho'_{r} > 0$} assign MF \textit{`m'} to FS \textit{`f'} and store \textit{`f'} in the \hspace*{2.1cm} array $ Y $, do $f=f+1$  \label{positive_rho_start}
			\If{new FS \textit{`f'} is not free}, deallocate all allocated FSs in path \hspace*{1.8cm} \textit{`p'}, empty the array $Y$,  $i=i+1$, Return  $ X=1,Y,i \, \text{and} \, f  $  \label{slot_not_available}
			\vspace{-0.2cm}
			\Else \, Return  $ X=2,Y,i \, \text{and} \, f  $  \label{slot_available}
			\EndIf  
			\ElsIf {$\rho'_{r}=0$} assign MF \textit{`m'} to FS \textit{`f'}, store  \textit{`f'} in  $Y$, Return  \hspace*{2.05cm} $ X=3,Y,i \, \text{and} \, f  $\label{zero_rho}
			\EndIf
			\Else \,\, Empty the array $Y$, $ i=i+1 $, Return  $ X= 1,Y,i \, \text{and} \, f  $  \label{slot_not_available2}
			\EndIf
			\Else \,\, Empty the array $Y$, $ i=i+1 $, Return  $ X= 1,Y,i \, \text{and} \, f  $  \label{slot_not_available3}
			\EndIf
		\end{algorithmic}	
	\end{algorithm}

	\subsubsection{Allocation for backup path} \label{back_allocation} 
	For each backup path $p_b \in  P_{r,p}^{b}$, allocation is done by calling the Subroutine 1  (line \ref{calling_sub_routine}, Algorithm \ref{working_algorithm}) which follows the similar approach used for working path. However, unlike working path, while finding maximum possible interference at targeted FS $\textit{`f'}$ in considered backup path `$p_b$', we only consider the failure of each link present in corresponding working path \textit{`p'} since `$p_b$' is enabled only when \textit{`p'} fails (lines \ref{urgent1}-\ref{urgent2}, Subroutine 1). Note that, Subroutine 1 returns the assigned FSs ($AFS_b$) if the allocation is successful in `$ p_b $' otherwise it returns empty array for $AFS_b$ (line \ref{return}, Subroutine 1). 
	
	\begin{algorithm} [h!]
		\floatname{algorithm}{} 
		\renewcommand{\thealgorithm}{} 
		\caption{Subroutine 1: BackupPath($\rho_{r},p,p_b,CFS^{p_b,p}_{r},Y$)} \label{backup_algorithm}
		\begin{algorithmic}[1]
			\State $i=1$
			\State $\rho'_{r}=\rho_{r}$  \label{rho_b}
			\State $f= CFS^{p_b,p}_{r}(i)$ 
			\For{failure of each link $e \in l_{p}$} \label{urgent1}
			\For{each link \textit{l} $\in l_{p_b,p}$} \label{path_allocation_b}
			\State Calculate IXT interference at \textit{`f'} using \eqref{constraint27} as,  $int^{f,e}_{l}$
			\EndFor
			\State Calculate end-to-end IXT interference at \textit{`f'} as,  $int^{f,e}_{p_b,p,r}=\underset{ l \in l_{p_b,p}}{\sum} int^{f,e}_{l} $ \label{int_bpath}
			\EndFor
			\State Calculate maximum possible IXT interference at \textit{`f'} as,  \hspace*{3.8cm}$int^{f,max}_{p_b,p,r}= \underset{e \in l_{p}}{max} \left(int^{f,e}_{p_b,p,r}\right)$ \label{urgent2}
			\State Calculate maximum possible IXT interference at \textit{`f'} in $p'$ and $ p'_b $ of $ R_{exi} $ considering the allocation of \textit{`f'} in `$ p_b $' as,  $int^{f,max}_{p',r'}$ \& $int^{f,max}_{p'_b,p',r'}$ $ \forall r' \in R_{exi} $
			\State Calculate ASE noise in \textit{`p'}  as $noi^{f}_{p,r}$ and in $p'$, $ p'_b $ of $ R_{exi} $ as $noi^{f}_{p',r'}$ \& $noi^{f}_{p'_b,p',r'}$ $ \forall r' \in R_{exi} $ using \eqref{variance}  \label{ASE_calculation2} 
			\State Compute worst case end-to-end SINR at \textit{`f'} in \textit{`p'} as $SINR^{f,min}_{p,r}$ and at \textit{`f'} in  $p'$, $ p'_b $ of $ R_{exi}$ as $SINR^{f,min}_{p',r'}$ \& $SINR^{f,min}_{p'_b,p',r'}$ 
			\State $SINR^{f,min}_{p,r} \leftarrow SINR^{f,min}_{p_b,p,r}$ , $p\leftarrow p_b$
			\State ($ X,Y,i,f $)=  \text{QoT}$(SINR^{f,min}_{p,r},SINR^{f,min}_{p',r'}, SINR^{f,min}_{p'_b,p',r'},m,f,i,p,\rho'_{r},Y) $
			\If{X=1} goto step \ref{rho_b}
			\ElsIf{X=2} goto step \ref{urgent1}
			\Else $ AFS_b \leftarrow Y $
			\EndIf
			\State Return $AFS_b$ \label{return}
		\end{algorithmic}	
	\end{algorithm} 
	
	
	%
	
	\subsubsection{Selection of working and backup paths}
	Once the allocation is done in considered working path \textit{`p'} and one of its backup paths `$p_b$', we find the value of objective function ($obj^{p_b}_{p,r}$, sum of highest indexed allocated FS in each link) by considering the spectrum allocations in \textit{`p'} and `$p_b$' including the FSs assigned in $ p' $ and $ p'_b $ of $ R_{exi} $ (line \ref{finding_objective}, Algorithm \ref{working_algorithm}). Next, we repeat the same process (described in previous subsection) for remaining backup paths ($p_b$) of working path \textit{`p'}. 
	Note that, for each \textit{`p'} and its corresponding `$p_b$' (if the allocation is successful in both paths), we compute the objective value $obj^{p_b}_{p,r}$ (line \ref{finding_objective}, Algorithm \ref{working_algorithm}). Thus, we get $K*K_b$ possible objective values for considered request \textit{`r'}. Now, we choose the minimum objective function owing to selected working path \textit{`p'} and its corresponding backup path `$p_b$' (lines \ref{min_objective_value}-\ref{min_objective_value2}, Algorithm \ref{working_algorithm}) and process the next request. 
	
	
	
	If we cannot allocate the required spectrum either in all working paths $p \in P_{r}^{w}$ or in all backup paths of each $p \in P_{r}^{w}$, we block the request and process the next request.
	

	\subsection{Complexity Analysis}
	The complexity of our proposed heuristic algorithm for static traffic scenario is $O(|C|^2|K||K_b||N||M||E|^2)$, where $|C|$, $|K|$, $|K_b|$, $|N|$, $|M|$ and $|E|$ are  total number of requests, working paths (of one request), backup paths (of one working path), FSs in a link, MFs and links, respectively. Next, complexity of heuristic for each incoming request under dynamic traffic scenario is $O(|C'||K||K_b||N||M||E|^2)$, where $ |C'| $ is the number of existing requests in the network. Complexity of our proposed sorting technique is $O(|C||K||K_b||E|)$.
	
	
	\section{MILP and heuristic for SRLG failure\label{SRLG}} 
	In this section, we extend our MILP and heuristic (proposed for QoT guarantee against single link failures) to design QoT guaranteed RSA in SBPP EON against SRLG failures. For this purpose, first, we calculate K shortest working paths for each request similar to link failure case. However, while calculating $K_b$ shortest backup paths for each working path, we make sure that all $K_b$ paths are SRLG disjoint (which is link disjoint for link failure case) to corresponding working path. Next, as explained in section \ref{introduction} and \ref{challenges}, similar to link failure case, different SRLG failures results in different $ R_w $ and $ R_b $ in the network and hence interference power on any FS is a function of failed SRLG. Therefore, robust optimization based RSA is required in this case also. By implementing similar concepts applied for link failure case (such as failing one SRLG at a time, finding worst case IXT interference considering each SRLG failure scenario etc.) proposed MILP and heuristic in sections \ref{MILP} and \ref{heuristic} can also be extended for SRLG failure by incorporating few changes as given below.

	    \hspace{-0.4cm}1) $ e $ denotes SRLG \\
		2) $ L $ denotes the set of SRLGs present in whole network \\
		3) $ L/l_p $ should be changed to $ L/SRLG_p $, where $ SRLG_p $ denotes the set of SRLGs present in path $ p $ \\
		4) $ e \in l_p $ should be changed to $ e \in SRLG_p $ \\
		5) Link disjoint constraint in section \ref{link_disjoint_constraint} should be changed to SRLG disjoint constraint which states that backup paths of two different requests can share the FSs if corresponding working paths are SRLG disjoint. For this purpose, variable \textit{l} in \eqref{constraint13} and \eqref{constraint17} should be changed to $ e $. Further, \eqref{constraint13} in its original form should also present as MILP constraint.\\
		6) Use the term SRLG failure in place of link failure.

	
	
	\section{Results and discussion} \label{results}
\begin{figure*} [t]
		\hspace{0.25cm}	
		\subfigure[]{\includegraphics[trim={2cm 10cm 21.5cm 0cm},height=0.4\textwidth, width=.3 \textwidth]{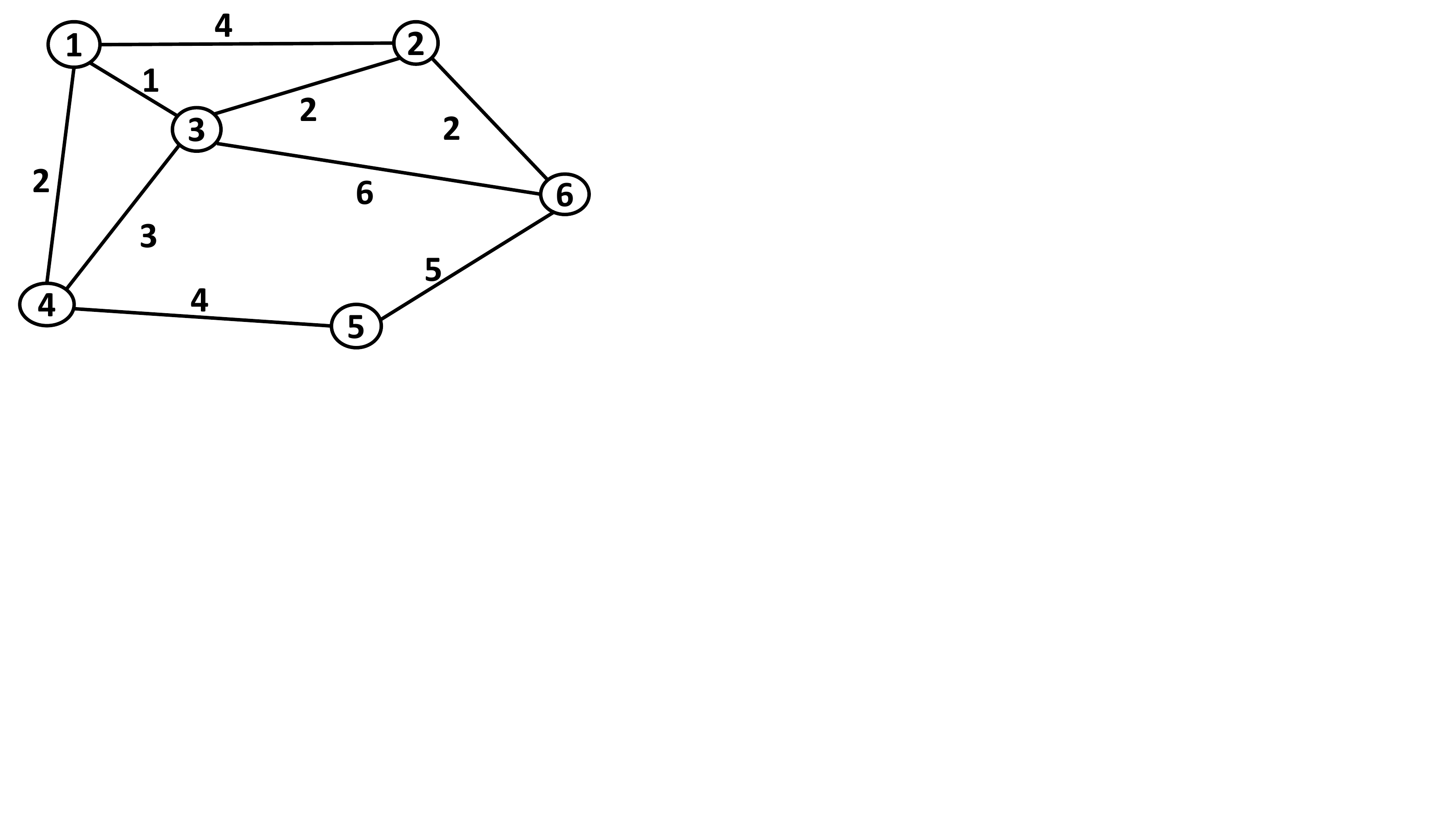} \label{6_node_network}}
		\hspace{2.2cm}
		\subfigure[]{\includegraphics[trim={5.5cm 6cm 3cm 7.5cm},height=0.55\textwidth, width=.4\textwidth]{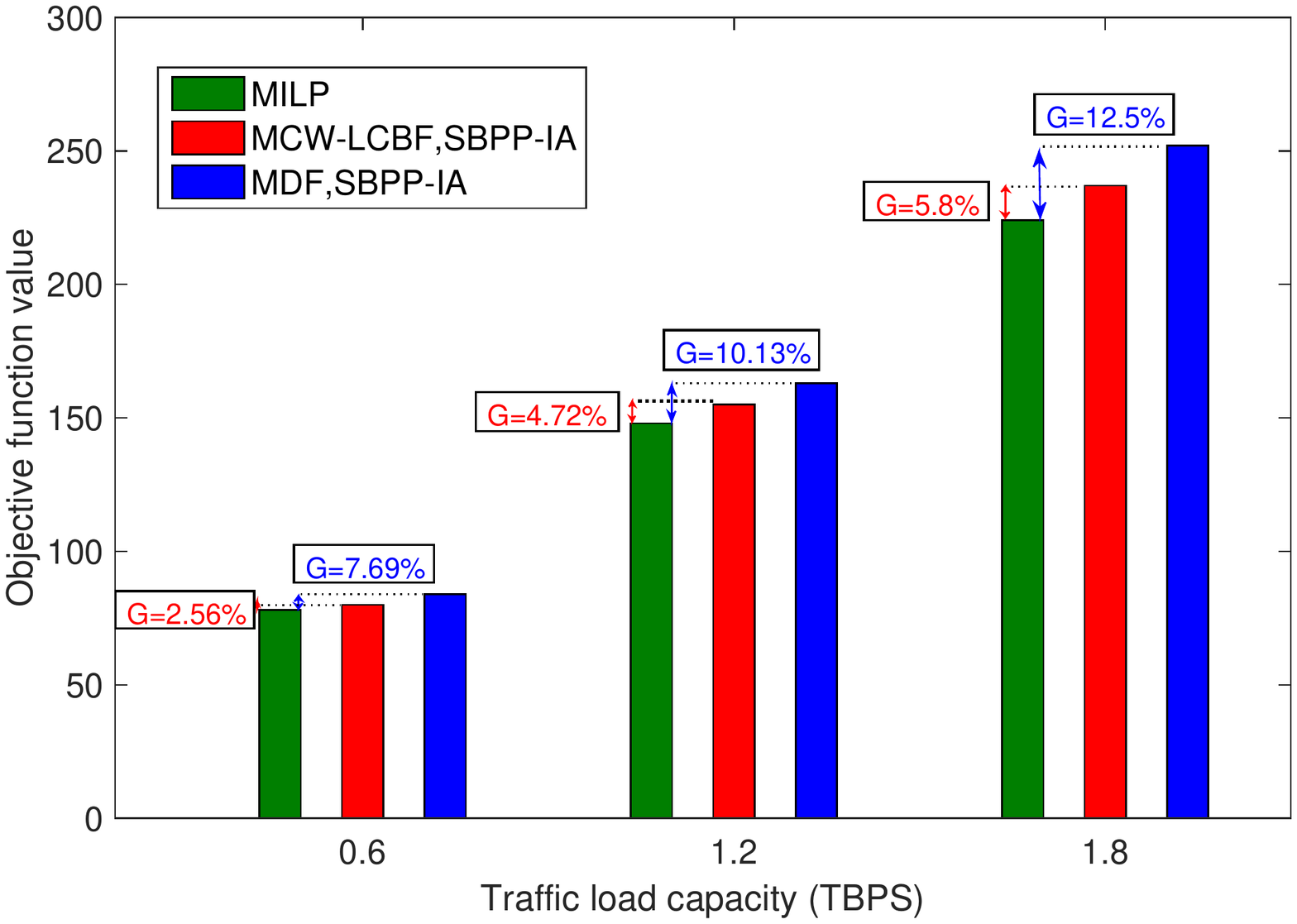} \label{load_vs_objective_MILP}}	
		\caption {(a)Sample 6-node network (link distance is in 100’s of kilometers).(b) Comparison of heuristic with MILP with respect to Optimality gap (G) for a 6-node network at $C_x$=-30 dB, N=30.}
		\label{results_static_traffic1}
	\end{figure*}

	In this section, we analyze the performance of the proposed MILP and the heuristic algorithm through simulation results. In particular, we evaluate the MILP under static traffic conditions whereas the heuristic is examined for both static and dynamic traffic scenarios. For this purpose, we adopt a 6-node network for MILP evaluation and a 14-node network for heuristics as shown in Fig. \ref{6_node_network} and \ref{14_node_network}, respectively. We assume that each link in considered networks is a bidirectional fiber and consists of 30 FSs in 6-node network and 350 FSs in 14-node network. Data rate of each request is chosen randomly following uniform distribution ranging from 10 to 70 Gbps for MILP and 10 to 700 Gbps for heuristics. Further, for each request, we compute $ K=3 $ working paths and for each working path, we calculate $ K_b=3 $ (wherever possible) backup paths which are link disjoint to corresponding working path. Next, we consider 4 different MFs namely BPSK, QPSK, 8-QAM and 16-QAM having approximate SINR thresholds 12.6 dB, 15.6 dB, 19.2 dB and 22.4 dB, respectively for a BER of $ 10^{-9} $ \cite{behera2019impairment}. Note that, for each value of total traffic load, we conduct the experiment 30 times and average the results. Simulation parameters are given in Table \ref{sim_par} for  broadcast and select (BS) node architecture \cite{behera2019impairment}.
	
%
%
	
	\vspace{-0.5cm}
	\subsection{Performance evaluation under static traffic scenario} 
	\subsubsection{For Small network} We compare the MILP with the proposed heuristic SBPP-IA for the 6-node network (Fig. \ref{6_node_network}) considering the existing and proposed sorting techniques MDF and MCW-LCBF, respectively in terms of optimality gap (\textit{G}) as shown in Fig. \ref{load_vs_objective_MILP}. Refer section \ref{sorting_requests} for the definition of \textit{G}.
	Fig. \ref{load_vs_objective_MILP} demonstrates that our SBPP-IA employing proposed MCW-LCBF outperforms the existing MDF.
	

	\vspace{0.3cm}
	\subsubsection{For Realistic network} 
	
	 \begin{figure}
	 	\centering 	
	 	\includegraphics[trim={5cm 7cm 18cm 0cm}, height=0.35\textwidth, width=0.25\textwidth]{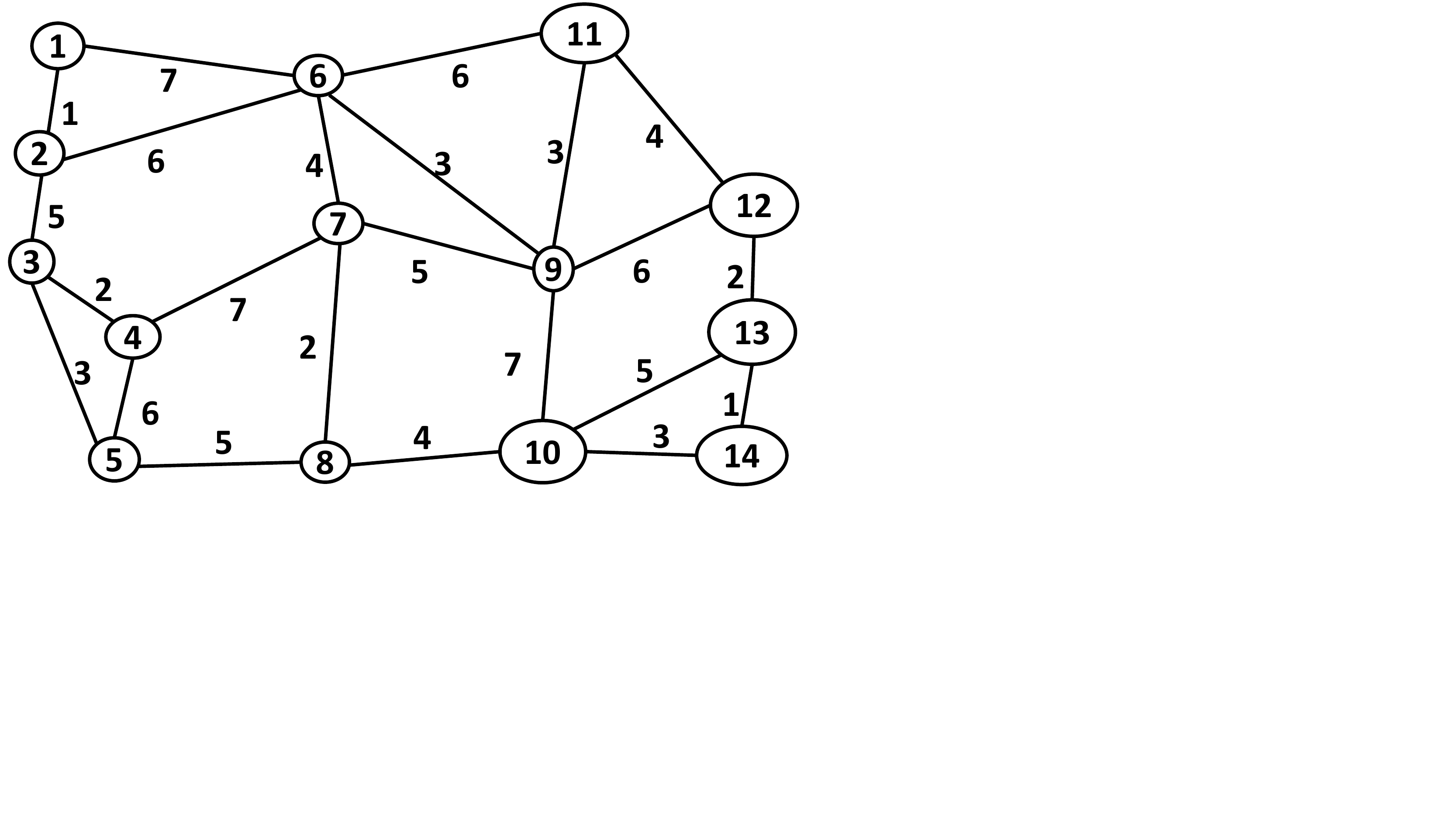}		
	 	\caption{14-node network (link distance is in 100’s of kilometers).}	
	 	\label{14_node_network}  
	 \end{figure}

\begin{figure*} [h]
			\centering
			\hspace{0.25cm}	
			\subfigure[]{\includegraphics[trim={5.5cm 7.7cm 4cm 8.9cm},height=0.4\textwidth, width=.35 \textwidth]{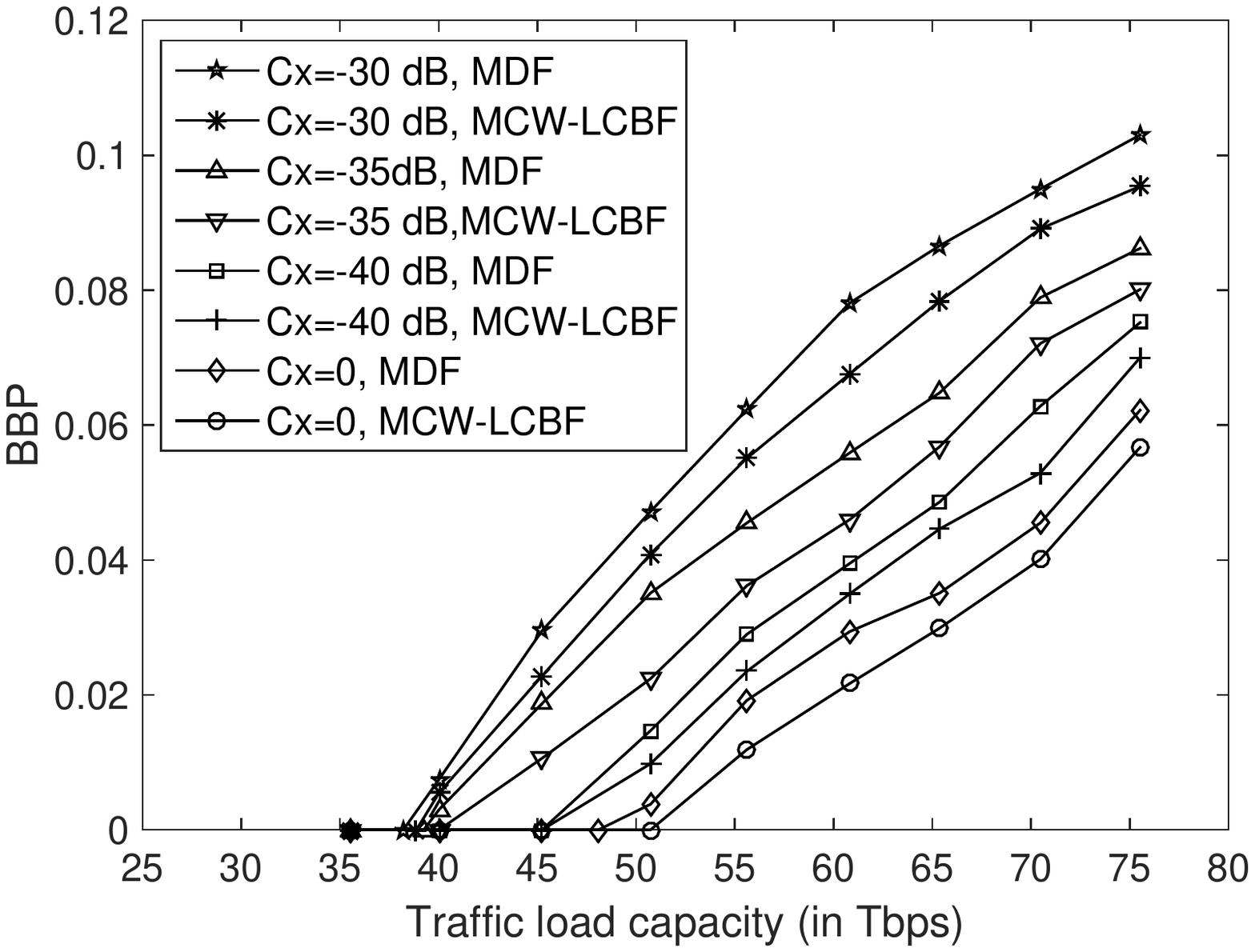} \label{traffic_vs_bbp_v7}}
			\hspace{1.3cm}
			\subfigure[]{\includegraphics[trim={4cm 6.8cm 3cm 8.5cm},height=0.398\textwidth, width=.36\textwidth]{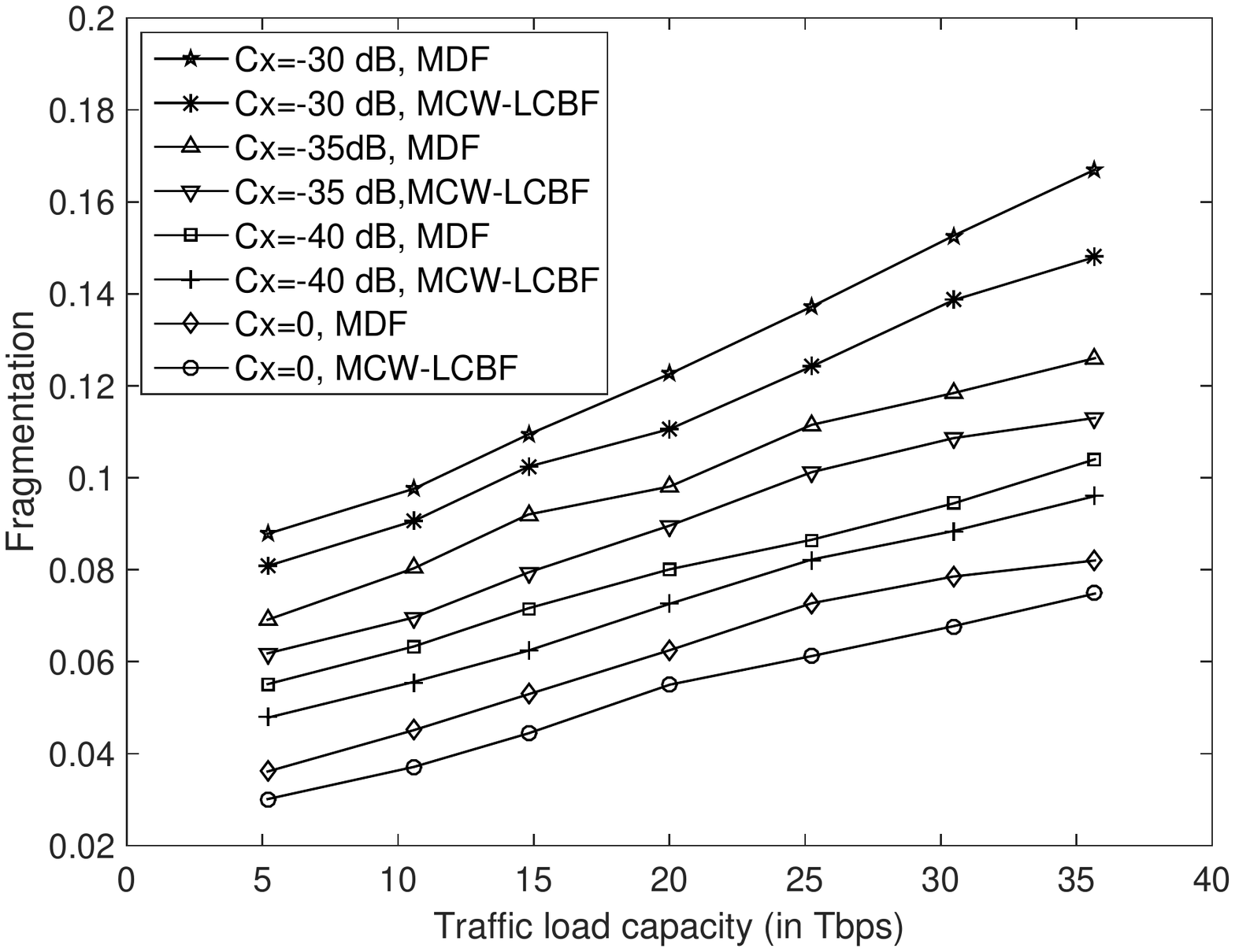} \label{traffic_vs_fragmentation_v3}}		
			\caption {Performance evaluation of proposed SBPP-IA for different crosstalk factors 
				($C_x $) under static traffic scenario for 14-node network. (a) Bandwidth blocking probability (b) Fragmentation.}
			\label{results_static_traffic}
		\end{figure*}

	As MILP involves complex computations for large scale simulations, we analyze the performance of the proposed SBPP-IA for the 14-node network (Fig. \ref{14_node_network}). In this regard, SBPP-IA employing MCW-LCBF and MDF are compared with respect to bandwidth blocking probability (BBP) and fragmentation as shown in \ref{traffic_vs_bbp_v7} and \ref{traffic_vs_fragmentation_v3}, respectively. BBP is calculated by finding the ratio of total bandwidth of blocked requests and overall bandwidth of all requests. Further, fragmentation is calculated as \cite{behera2018effect}:
	
	\vspace{-0.3cm}
	\begin{align}
		\hspace{-0.5cm}
		F_r=1-\frac{\text{largest continuous free FSs block}}{\text{total free FSs}}
	\end{align}

	
	As shown in Fig. \ref{results_static_traffic}, simulations are carried out for different values of crosstalk factor ($ C_x $). Fig. \ref{traffic_vs_bbp_v7} describes that BBP increases as traffic load increases. This is due to the effect of higher interference and insufficient resources with the increase in traffic load. Further, higher $ C_x $ value introduces higher interference which eventually increases the BBP at each load.
	\vspace{0.2cm}
	
	When it comes to fragmentation, as BBP starts from 40 Tbps (Fig. \ref{traffic_vs_bbp_v7}), we evaluate the performance of SBPP-IA in terms of fragmentation for the overall load of 5-35 Tbps as shown in Fig. \ref{traffic_vs_fragmentation_v3}. We observe that, similar to BBP, fragmentation also increases as traffic load and $C_x$ increase. The reason is that, the increase in traffic load causes more interaction between the FSs of different requests introducing more interference at FSs and therefore not allowing any MF to be allocated. In that case, to minimize the effect of interference, requests are more fragmented during the allocation increasing overall fragmentation of the network.
	
	It is interesting to observe that, our SBPP-IA employing the proposed sorting technique MCW-LCBF results in better performance in terms of BBP and fragmentation than the existing MDF technique (refer Fig. \ref{traffic_vs_bbp_v7} and \ref{traffic_vs_fragmentation_v3}). This is because, MDF sorts the requests only based on the data rate demand of each request. On the other hand, our proposed MCW-LCBF considers the probability of congestion in each working and backup path of every request and sorts the requests in such a way that there exists maximum benefit in the allocations of both working and backup paths as explained in section \ref{sorting_requests}. 
	\begin{table}[h!]
		\centering
		\caption{SIMULATION PARAMETERS}
		\vspace{-0 in}
		\label{sim_par}
		\begin{tabular}{|l|l|}
			\hline
			Parameter                                  & Value     \\ \hline
			LO power ($P_{lo}$), Received power ($P_r$)                 & 0 dBm, -12 dBm    \\ \hline
			Responsivity ($R_a$),  Operating frequency ($f_c$)                       & 0.7 A/W,  193.1 THz     \\ \hline
			Spontaneous Emission factor ($n_{sp}$)            & 2        \\ \hline
			Switch through loss (dB) assuming                        & 3 $\ceil{log_{2}Q}+L_{WSS}$, \\
			broadcast and select architecture \cite{behera2019impairment}	& $Q$: inputs/outputs  	 
			\\ \hline
			Fiber attenuation ($\alpha$), WSS loss $(L_{WSS})$                       & 0.2 dB/km, 2 dB \\ \hline
			Tap loss ($L_{tap}$) & 1 dB \\ \hline
			
			EDFA spacing $(E_s)$  \cite{zhao2015nonlinear}                           & 100 km   \\ \hline
			Input EDFA gain ($G_{in}$)   & 22 dB \\ \hline
			Output EDFA gain ($G_{out}$) for 6-node                              & 5 dB at node 3,\\
			(see Fig. \ref{6_node_network})   &8 dB elsewhere\\ \hline             
			Output EDFA gain ($G_{out}$) for 14-node                               & 5 dB at nodes 1, 14 \\ 
			(see Fig. \ref{14_node_network})  & 11 dB at nodes 6, 9     \\  
			& 8 dB elsewhere 
			\\ \hline				
			Crosstalk factor ($Cx$)        &[-30, -35 -40] dB  \\ \hline
			Electrical bandwidth ($B_e$)      & 7 GHz      \\ \hline
			Planck's constant (h)  & $6.62 \times 10^{-34}$ J.s \\ \hline
			
		\end{tabular}
	\end{table}
	
	\subsection{Performance evaluation under dynamic traffic scenario}
	For dynamic traffic scenario, we assume that each request's arrival follows a Poisson process while holding times are negative exponentially distributed \cite{wang2015distance}. Here, we simulate 1,00,000 requests where the data rate demand of each request is chosen between 10-700 Gbps obeying uniform distribution. Algorithm \ref{working_algorithm} (excluding the sorting of requests, step 1) proposed in section \ref{heuristic_algorithm} is used for the allocation of dynamic traffic. Note that, after the departure of any request, allocated FSs in working and backup path of departed request are released which can be used to accommodate the incoming requests. However, we do not release the backup FSs of departed request involved in sharing with existing requests.

	Here, we compare the performance of our heuristic SBPP-IA with SBPP-spectrum window plane (SBPP-SWP) algorithm proposed in \cite{wang2015distance} in which RSA was designed for SBPP EON without considering the effect of PLIs. Noteworthy, after the design of SBPP-SWP algorithm in \cite{wang2015distance}, various authors extended SBPP EON to different areas such as IP over EON, traffic grooming, multicast RSA etc. Therefore, to compare the performance of our proposed SBPP-IA algorithm, we use the standard algorithm (SBPP-SWP) proposed for SBPP EON in \cite{wang2015distance} as a benchmark. For this purpose, we simulate SBPP-IA and SBPP-SWP algorithms for the 14-node network (Fig. \ref{14_node_network}) under dynamic traffic scenario. Further, since QoT guaranteed allocation is imperative for any optical network, above mentioned SBPP EON extended works implemented in different areas (IP over EON etc.) should be readdressed considering PLIs by using robust optimization as demonstrated in our MILP and heuristic algorithm.
		\begin{figure*} 
			\centering
			\subfigure[]{\includegraphics[trim={2.9cm 8.2cm 1.8cm 8.65cm}, height=0.4\textwidth, width=0.47\textwidth]{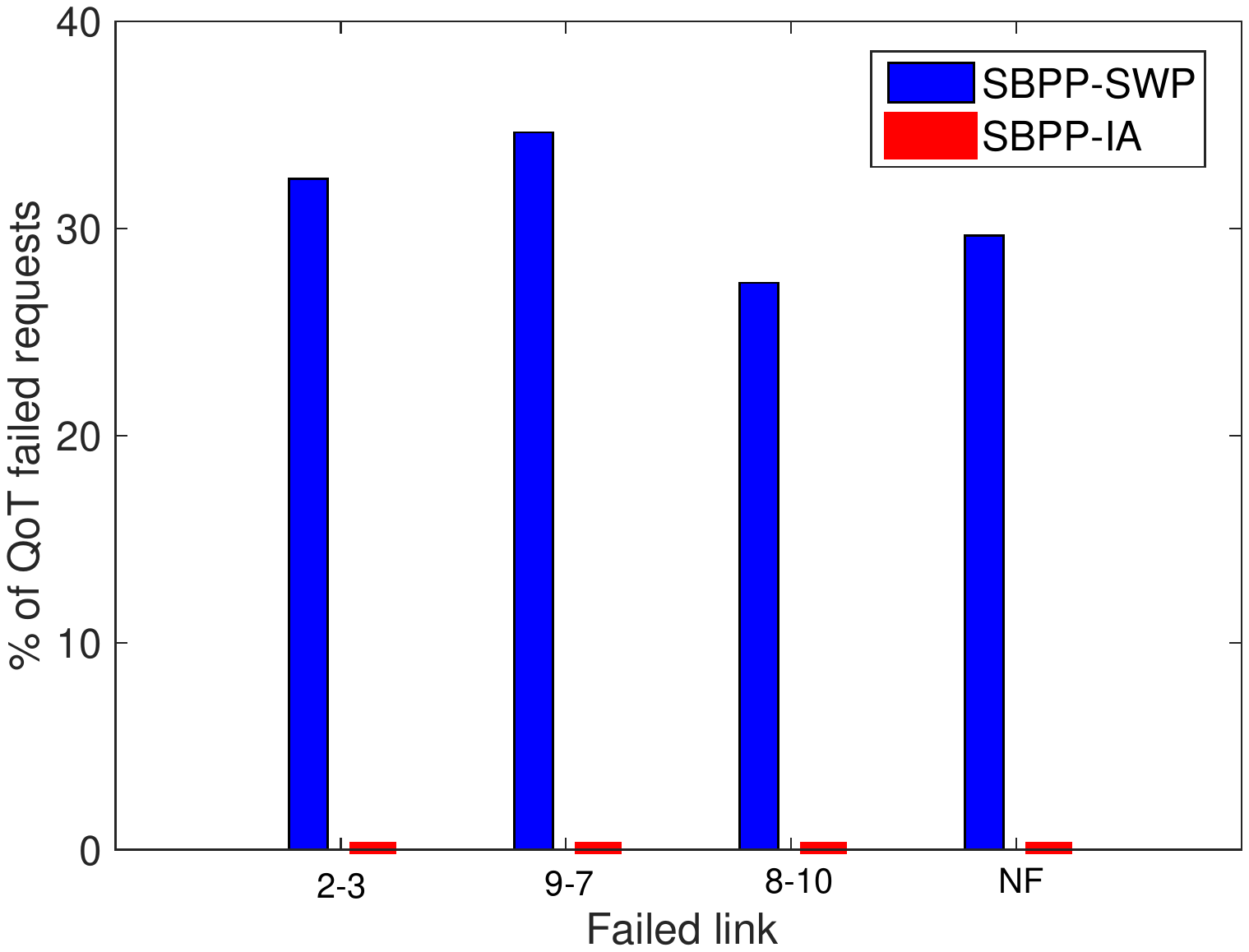}
				\label{load_vs_percentofrequestsfailed_bar_linkfail_nofail_v1}}
			\subfigure[]{\includegraphics[trim={2cm 7.9cm 2.4cm 9cm}, height=0.385\textwidth, width=0.47\textwidth]{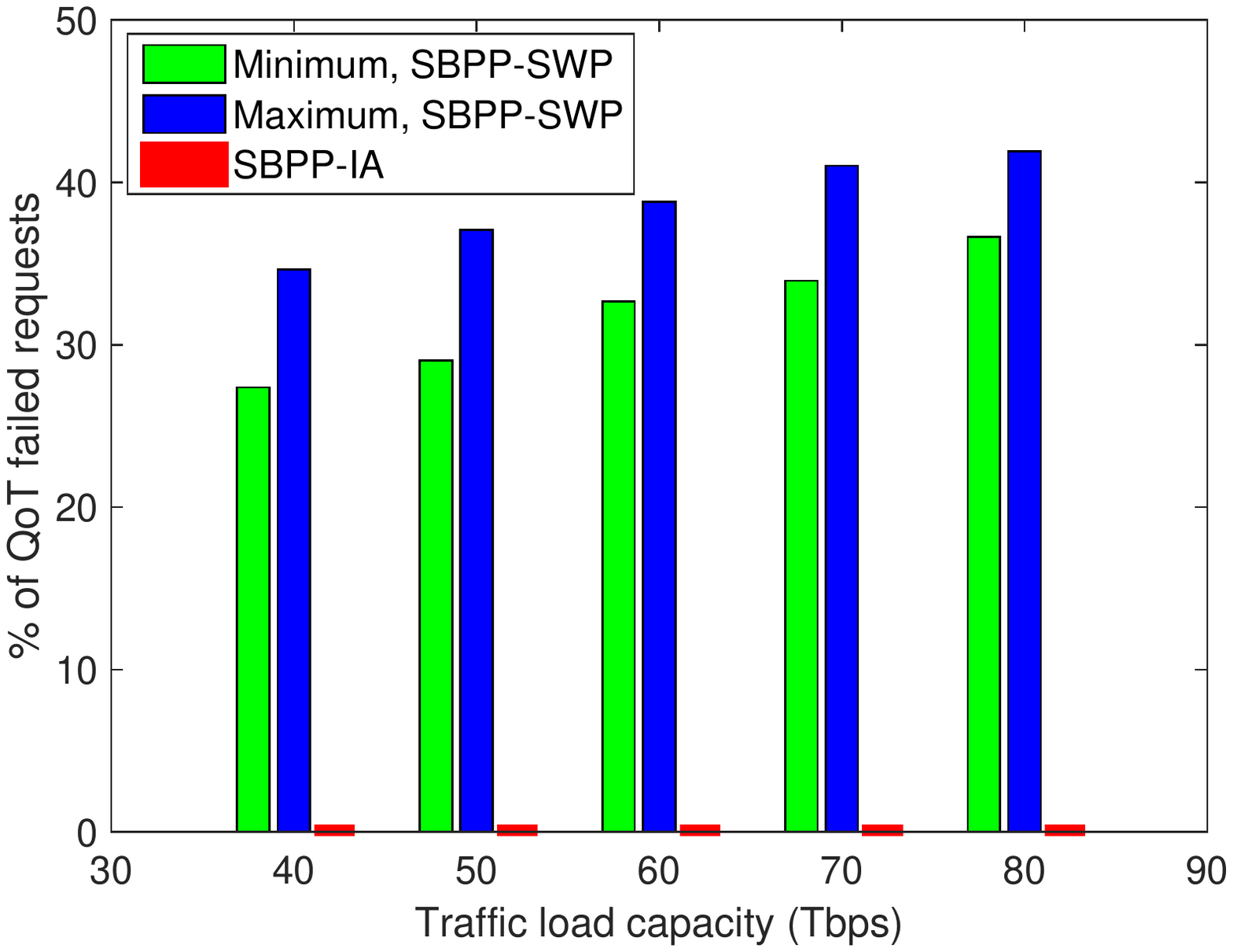}
				\label{load_vs_percentofrequestsfailed_bar_worstcase_v2}}
			\caption {Performance comparison of proposed SBPP-IA with SBPP-SWP under dynamic traffic scenario for 14-node network at $ C_x$=-30 dB. (a) Percentage of  QoT failed requests for SBPP-SWP under different link failure scenarios at 40 Tbps network load (b) Maximum and minimum possible percentage of QoT failed requests for SBPP-SWP.}
			\label{results_dynamimc_traffic1}
			\vspace{0.2cm}
		\end{figure*}
	
	
	
	
	\vspace{0.05cm}
	First, we evaluate performance comparison between SBPP-IA and SBPP-SWP in terms of percentage of QoT failed requests. To calculate the percentage of QoT failed requests for SBPP-SWP, we simulate this algorithm and perform RSA for working and backup path of each request. Next, for a given total traffic load value, by failing one link in the network, we compute the percentage of QoT failed requests while considering PLIs. Note that, if either active working path or active backup path of any request is failed to meet minimum QoT requirements then the corresponding request is considered to be QoT failed request.
	We repeat this procedure for each single link failure and no link failure cases and corresponding results are plotted in Fig. \ref {load_vs_percentofrequestsfailed_bar_linkfail_nofail_v1}. Due to space constraint, we have shown results correspond to few single link failure cases and no link failure (denoted by NF on X-axis, Fig. \ref {load_vs_percentofrequestsfailed_bar_linkfail_nofail_v1}) case for a network load of 40 Tbps (Fig. \ref {load_vs_percentofrequestsfailed_bar_linkfail_nofail_v1}). 
	
	
	
	\vspace{0.1cm}
	It is interesting to observe that, each single link failure results in different percentage of QoT failed requests which is again different for no link failure case as illustrated in Fig. \ref{load_vs_percentofrequestsfailed_bar_linkfail_nofail_v1}.
	Therefore, at each traffic load, by considering each single link failure and no link failure conditions, we calculate the minimum and maximum possible percentage of QoT failed requests denoted by `Minimum' and `Maximum,' respectively as shown in Fig. \ref{load_vs_percentofrequestsfailed_bar_worstcase_v2}. Finally, results in Fig. \ref{load_vs_percentofrequestsfailed_bar_worstcase_v2} summarize that significant amount of requests are failing to meet minimum QoT requirements at each traffic load if the allocation is done using SBPP-SWP algorithm in which the effect of PLIs are not considered. On the other hand,
	we observe that, our SBPP-IA algorithm guarantees minimum QoT assured allocation in both working and backup paths for each request under any single link failure and no link failure conditions at any traffic load.
	
		\begin{figure*}
			\centering
					\hspace{0.3cm}
			\subfigure[]{\includegraphics[trim={4cm 6.9cm 3.5cm 10cm},height=0.4\textwidth, width=.44\textwidth]{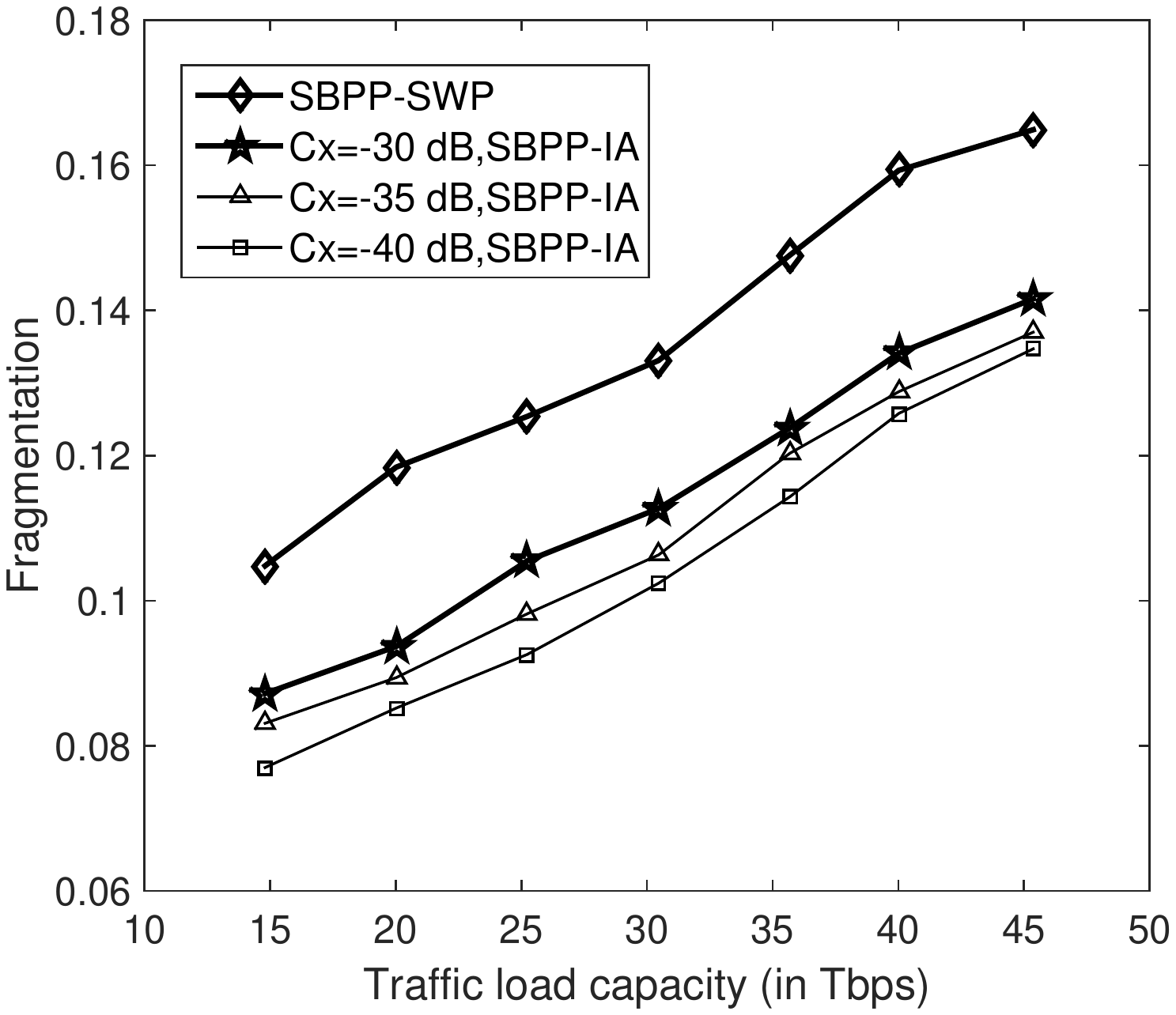} \label{traffic_vs_fragmentation_comparison_v1}} 
			\hspace{-0.1cm}
			\subfigure[]{\includegraphics[trim={3cm 7.4cm 3.5cm 9cm},height=0.45\textwidth, width=.49\textwidth]{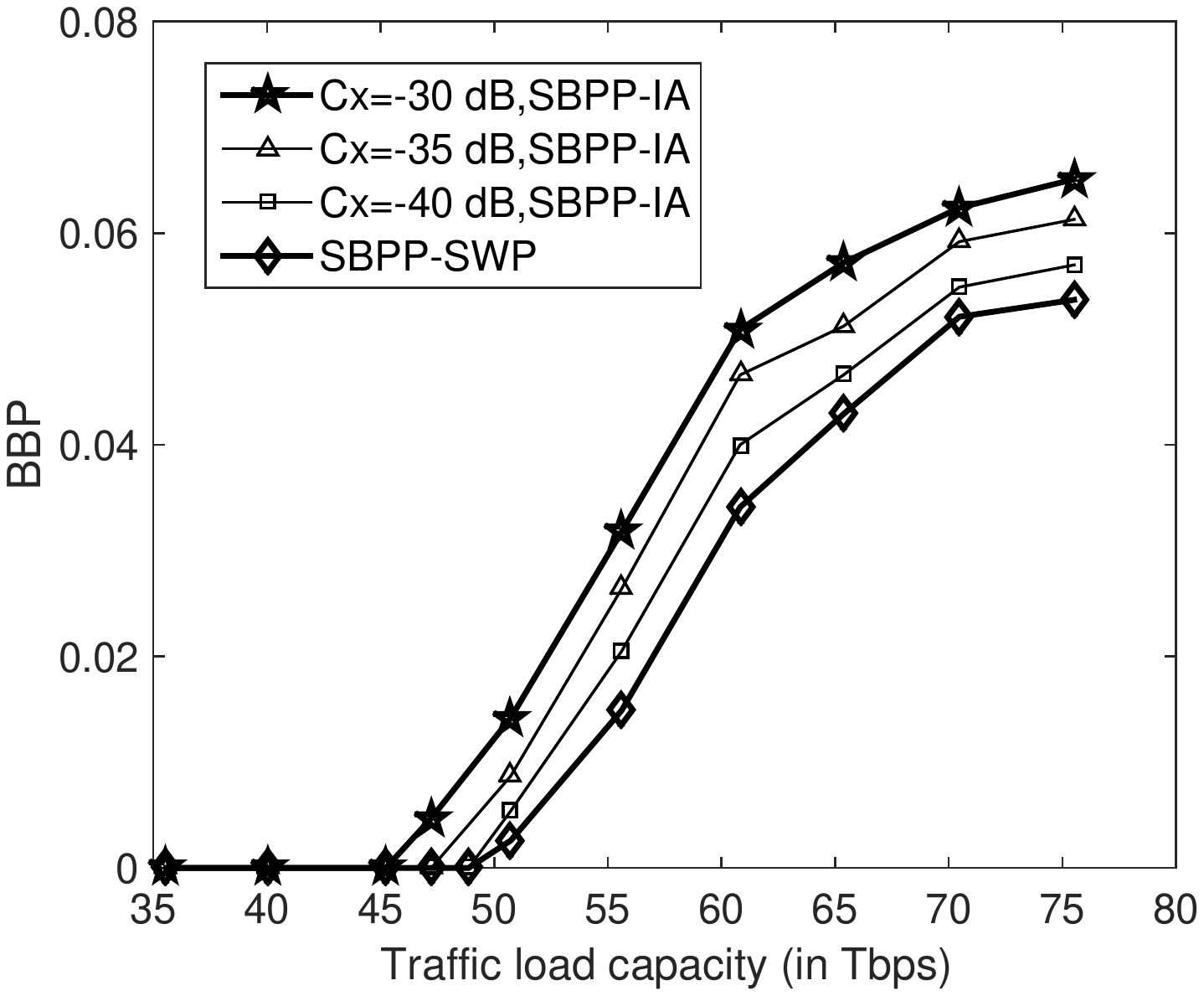}
				\label{traffic_vs_bbp_comparison_v1}} 	
			\caption {Performance comparison of proposed SBPP-IA with SBPP-SWP under dynamic traffic scenario for 14-node network. (a) Fragmentation (b) BBP    
				\label{results_dynamimc_traffic2}}	
		\end{figure*}

\begin{figure*}
			\centering
			\hspace{-1.4cm}	
			\subfigure[]{\includegraphics[trim={3cm 7.6cm 2cm 9.1cm},height=0.45\textwidth, width=.53\textwidth]{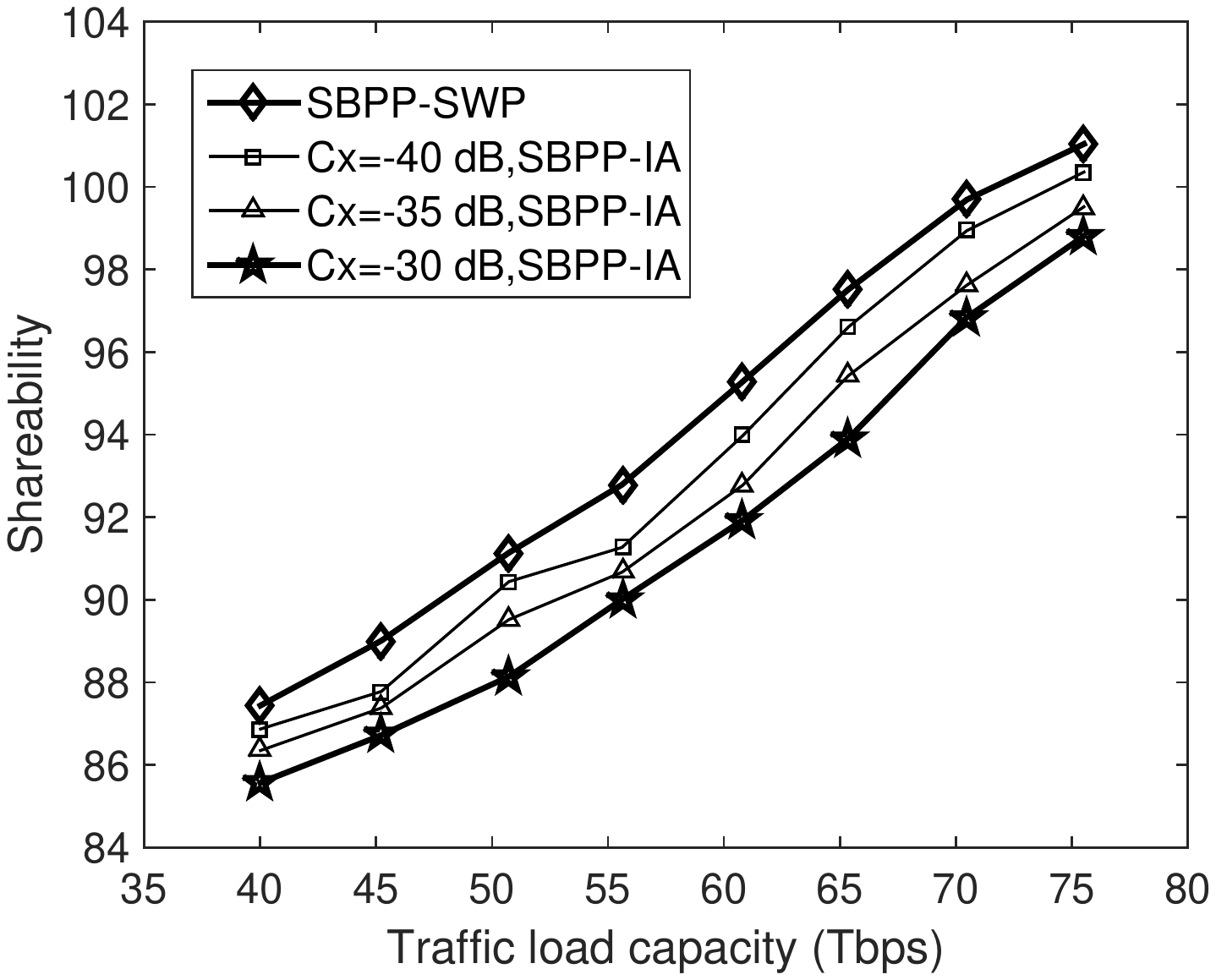}		\label{traffic_vs_percentofslotssharing_comparison_v2}}
			\subfigure[]{\includegraphics[trim={2.5cm 6.8cm 2.1cm 9cm},height=0.44\textwidth, width=.5\textwidth]{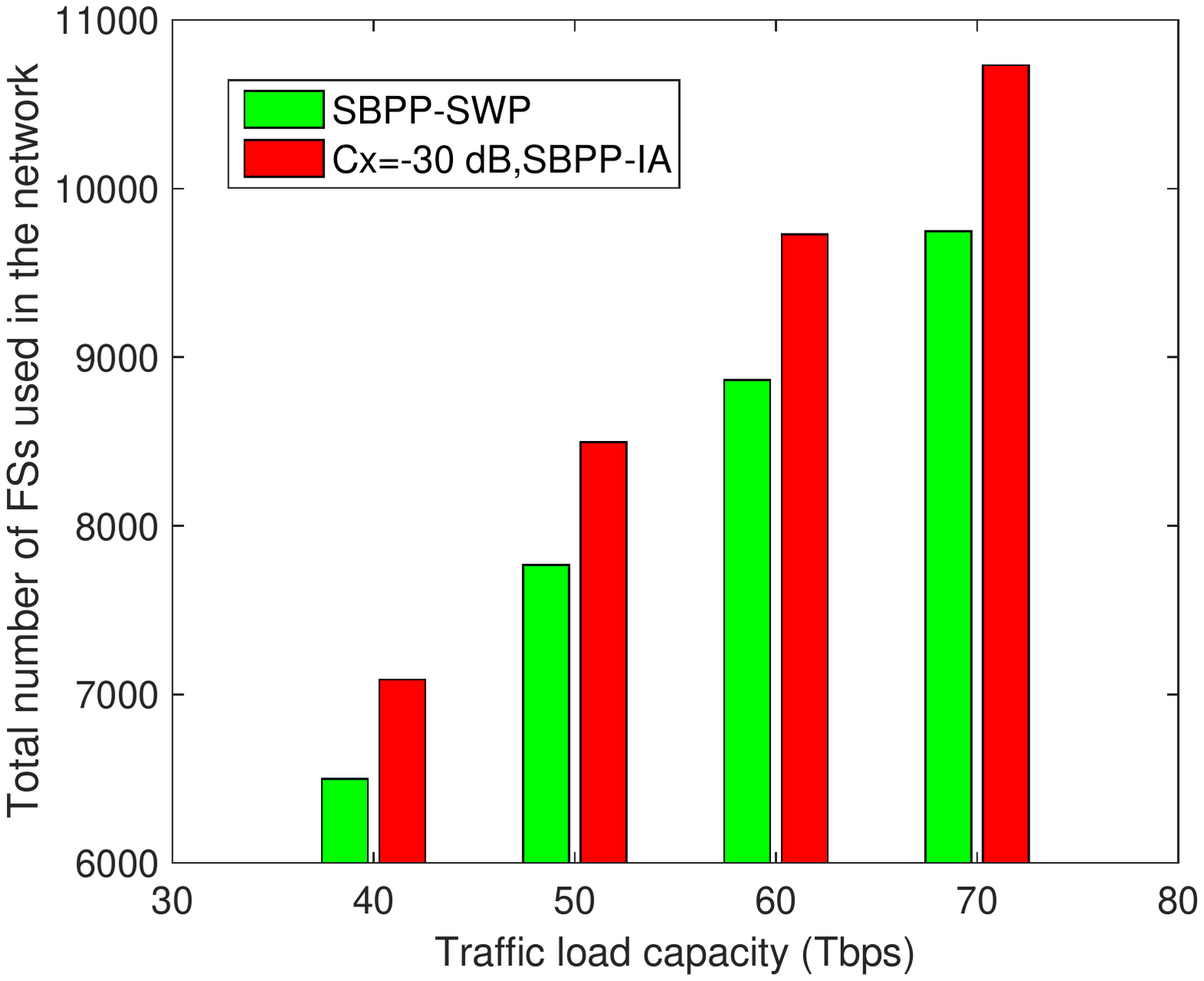}	\label{load_vs_totalslots_comparison_v1}} 
			\hspace{-1.8cm}
			\caption {Performance comparison of proposed SBPP-IA with SBPP-SWP under dynamic traffic scenario for 14-node network. (a) Shareability (b) Total number of FSs used. 
				\label{results_dynamimc_traffic3}}
		\end{figure*}

	\vspace{0.1cm}
	Next, Fig. \ref{traffic_vs_fragmentation_comparison_v1} and \ref{traffic_vs_bbp_comparison_v1} describe that fragmentation and BBP increase with increase in load for SBPP-IA and SBPP-SWP algorithms. Further, performances in terms of BBP and fragmentation are improved in SBPP-IA as $ C_x $ reduces. 
	When it comes to backup sharing, it is to be noted that, our proposed SBPP-IA attempts to minimize the fragmentation in each link which in turn offers maximum possible sharing among the backup FSs. To demonstrate this, we calculate the shareability at various traffic loads for different $ C_x $ values and the corresponding results are shown in Fig. \ref{traffic_vs_percentofslotssharing_comparison_v2}. Shareability denotes the percentage of backup FSs involved in sharing among total allocated backup FSs and is calculated as follows:\\
	
	\vspace{-1cm}
	
	\begin{align}
		\text{Shareability}=  \frac{\sum_{l} \sum_{f} (s_{f,l}-1)}{N_b} \times 100
	\end{align}

	\hspace{-0.4cm}$ \text{Where,} \, s_{f,l}: \text{Number of requests shared the FS \textit{`f'} in link \textit{`l',}} \\ \hspace*{1cm} N_b:\text{Number of backup FSs allocated in the network.}$\\
	
	\vspace{-0.2cm}
	
	Results in Fig. \ref{traffic_vs_percentofslotssharing_comparison_v2} demonstrate that, shareability increases as traffic load grows since opportunities for sharing the backup FSs among requests get increased with increase in traffic load. 
	Further, effect of crosstalk is also observed in Fig. \ref{traffic_vs_percentofslotssharing_comparison_v2} where backup shareability improves as $ C_x $ decreases. This is because, requests can be closely allocated in presence of less interference which increases the backup sharing. It is evident from Fig. \ref{traffic_vs_percentofslotssharing_comparison_v2} that our objective function achieves significant backup shareability in the network. In addition, shareability of SBPP-SWP algorithm at different loads is also presented in Fig. \ref{traffic_vs_percentofslotssharing_comparison_v2}. Next, Fig. \ref{load_vs_totalslots_comparison_v1} depicts that, total number of FSs used in the network increases with increase in traffic load both in case of SBPP-IA and SBPP-SWP
	algorithms.
	\vspace{0.1cm}
	Now, we illustrate the performance comparison between SBPP-IA and SBPP-SWP in terms of fragmentation, BBP, shareability and total number of FSs used while considering $C_x=-30$ dB for SBPP-IA. For better understanding, results corresponding to SBPP-IA ($C_x=-30$ dB) and SBPP-SWP are represented with bold lines as in Fig. \ref{traffic_vs_fragmentation_comparison_v1}, \ref{traffic_vs_bbp_comparison_v1} and \ref{traffic_vs_percentofslotssharing_comparison_v2}. It is observed that, SBPP-SWP shows better performance than our SBPP-IA algorithm in terms of BBP, shareability and total number of FSs used (refer Fig. \ref{traffic_vs_bbp_comparison_v1} and \ref{results_dynamimc_traffic3}) as our algorithm considers the effect of PLIs during the allocation.
	Note that, this performance difference is observed to be insignificant due to the following reasons. Firstly, we employ the efficient spectrum allocation technique named bitloading which is demonstrated in \cite{behera2019impairment}.
	Secondly, our objective function minimizes fragmentation in the network. On the other hand, our SBPP-IA outperforms the SBPP-SWP in terms of percentage of QoT failed requests and fragmentation as shown in Fig. \ref{results_dynamimc_traffic1} and \ref{traffic_vs_fragmentation_comparison_v1}, respectively. 
	
	\vspace{0.1cm}
	For better understanding, we observe our results at 70 Tbps total traffic load (Fig. \ref{results_dynamimc_traffic1},\ref{results_dynamimc_traffic2} and \ref{results_dynamimc_traffic3}) to compare SBPP-IA and SBPP-SWP. At 70 Tbps, SBPP-SWP performs slightly better compared to SBPP-IA by a percentage of 1.03, 2.85 6.12 in terms of BBP, shareability and total number of FSs used, respectively. Conversely, our SBPP-IA outperforms SBPP-SWP by a percentage of 33.94-41.02 (depending upon location of link failure) and 2.33 in terms of QoT failed requests and fragmentation, respectively.

	%
	%
	
	
	\subsection{Performance evaluation of heuristic for SRLG failure case}
		\begin{figure} [h]
			\hspace{2cm}
			\includegraphics[trim={1.5cm 2cm 7cm 6.8cm}, height=0.66 \textwidth, width=0.45 \textwidth]{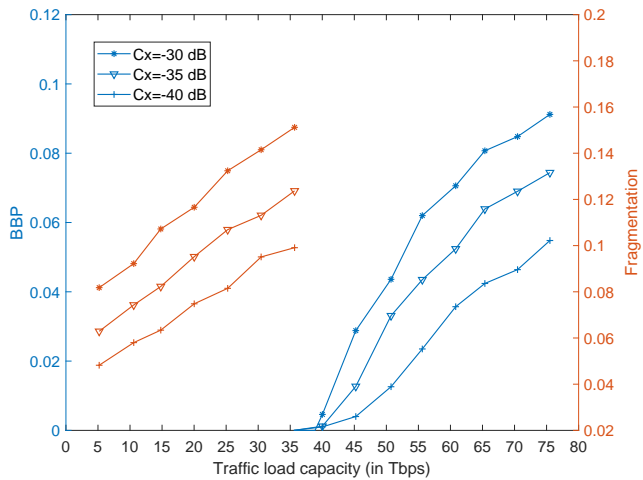}		
			\vspace{-1.3cm}
			\caption{Performance of proposed SBPP-IA using MCW-LCBF sorting technique against node failure scenario under static traffic for 14-node network in terms of BBP and fragmentation.}	
			\label{results_node_failure}     
		\end{figure}
	In this section, we examine the proposed heuristic (refer section \ref{SRLG}) against SRLG failure for static traffic scenario. Note that, we simulated the heuristic and generated results considering the network node as SRLG for the 14-node network shown in \ref{14_node_network}. Further, for each request, we calculate $ K=3 $ working paths and for each working path, we compute $ K_b=3 $ (wherever possible) backup paths which are node disjoint to corresponding working path. 
		If there exists no node disjoint paths for each working path of a request, we block that particular request. We evaluate the performance of heuristic in terms of BBP and fragmentation as shown in Fig. \ref{results_node_failure}. As can be seen in Fig. \ref{results_node_failure}, graphs follow similar pattern as in Fig. \ref{results_static_traffic} with the variation of load and $ C_x $ due to the same reason described for earlier results.


	\section{Conclusion} \label{conclusion}
	In this paper, we focus on the design of SBPP based EON which provides a minimum QoT assured RSA against PLIs under any single link/SRLG failure for static and dynamic traffic scenarios. In this regard, we considered PLIs such as in-band crosstalk (IXT) along with ASE and beating noise terms in our RSA design. Next, we formulated an MILP for smaller networks and proposed MCW-LCBF sorting technique followed by the SBPP-IA heuristic for larger networks under static traffic scenario. Results demonstrated that our MCW-LCBF outperforms widely used existing MDF in terms of optimality gap, BBP and fragmentation. Next, we compared our proposed SBPP-IA algorithm with the existing SBBP-SWP algorithm under dynamic traffic scenario. It is evident from simulation results that, though the minimum QoT assured resource allocation is an important attribute of optical network, existing SBPP-SWP results in significant QoT failed requests whereas our proposed SBPP-IA guarantees 100\% QoT assured allocation in working and backup paths under any single link failure and no link failure conditions by compromising a little in terms of BBP, shareability and total number of FSs used. 
	
	
	
	
	
	\appendix
	\section{Linearizations}
	%
	
	1) Logical AND: $  Z=XY $ can be linearized as given below \\
	\hspace*{0.65cm}
	$Z \geq 0 ;
	Z \leq X ;
	Z \leq Y ;
	Z \geq X+Y-1 ; $ \\
	\hspace*{0.35cm}2) Maximum of continuous variables: \\
	\hspace*{0.85cm}$ Y=max(X_1,X_2,...,X_n) $ can be linearized as follows:
	\hspace*{0.85cm}$ L_i \leq X_i \leq U_i \, ; \, \forall i = \left\lbrace 1, 2, \dots,  n \right\rbrace  $ \\ 
	\hspace*{0.85cm}$ Y \geq X_i \, ; \, \forall i $\\
	\hspace*{0.85cm}$ Y \leq X_i + (U_{max}-L_i)(1-D_i) \, ; \, \forall i $  \\
	\hspace*{0.85cm}$  \sum_{i=1}^{n}D_i =1 \, ;$ \\
	\hspace*{0.9cm}Where, $ L_i $ and $ U_i $ are lower and upper bounds  of \hspace*{0.9cm}$ X_i $,  respectively,
	$ U_{max}= max(U_1,U_2....U_n)$.  \textcolor{blue}{In our}\\
	\textcolor{blue}{\hspace*{0.87cm} case, $ L_i =0, U_i = \ceil*{\dfrac{N_d-2}{2}}\times N_{xc} \times P_r \times C_x, \forall i $ \\  
		\hspace*{0.84cm} Where, $ N_d $= maximum nodal degree in the network, \\
		\hspace*{0.86cm} $ N_{xc} $= total number of nodes in the network.}
	
	\hspace*{0.66cm}Further, we define $ D_1,D_2,...,D_n $ as binary variables \hspace*{0.82cm} for which $ D_i=1$ if $X_i $ is the maximum value, 0 \hspace*{0.89cm} otherwise.

\bibliography{references}

\end{document}